%% file: main.tex
\documentclass[journal]{IEEEtran}
\usepackage{authblk}
\usepackage[utf8]{inputenc}
\usepackage[T1]{fontenc}
\usepackage[colorlinks,linkcolor=blue,citecolor=blue]{hyperref}
\usepackage[dvipsnames,table]{xcolor}
\usepackage{xcolor}
\usepackage{booktabs}
\usepackage[framemethod=tikz]{mdframed}
\usepackage{algorithm}
\usepackage{adjustbox}
\usepackage{balance}
\usepackage[inline]{enumitem}
\usepackage{algorithmic}
\usepackage{graphicx}
\usepackage{caption}
\usepackage{hhline}
\usepackage{tabularx}
\usepackage[scaled]{beramono}

\newcolumntype{R}{>{\raggedleft\arraybackslash}X}

\usepackage{xspace}
\usepackage{pgfplots}
\pgfplotsset{compat=1.15}
\usetikzlibrary{patterns}
\usepackage{listings}

\definecolor{light-gray}{gray}{0.95}
\definecolor{pgreen}{RGB}{5,205,107}
\definecolor{pblue}{RGB}{2,154,223}

\makeatletter
\lst@AddToHook{OnEmptyLine}{\addtocounter{lstnumber}{-1}}
\makeatother

\makeatletter
\newenvironment{btHighlight}[1][]
{\begingroup\tikzset{bt@Highlight@par/.style={#1}}\begin{lrbox}{\@tempboxa}}
{\end{lrbox}\bt@HL@box[bt@Highlight@par]{\@tempboxa}\endgroup}

\newcommand\btHL[1][]{%
  \begin{btHighlight}[#1]\bgroup\aftergroup\bt@HL@endenv%
}
\def\bt@HL@endenv{%
  \end{btHighlight}%
  \egroup
}
\newcommand{\bt@HL@box}[2][]{%
  \tikz[#1]{%
    \pgfpathrectangle{\pgfpoint{1pt}{0pt}}{\pgfpoint{\wd #2}{\ht #2}}%
    \pgfusepath{use as bounding box}%
    \node[anchor=base west, fill=orange!30,outer sep=0pt,inner xsep=1pt, inner ysep=0pt, rounded corners=3pt, minimum height=\ht\strutbox+1pt,#1]{\raisebox{1pt}{\strut}\strut\usebox{#2}};
  }%
}
\makeatother

\lstdefinestyle{bytecode}{
    language={SQL},basicstyle=\ttfamily\scriptsize, 
    moredelim=**[is][\btHL]{`}{`},
    moredelim=**[is][{\btHL[fill=green!30]}]{@}{@},
}

\usepackage{amsthm}
\theoremstyle{definition}

\usepackage[framemethod=tikz]{mdframed}
\mdfdefinestyle{mpdframe}{
    frametitlebackgroundcolor   =black!15,
    frametitlerule              =true,
    roundcorner                 =1pt,
    middlelinewidth             =1pt,
    innermargin                 =0.1cm,
    outermargin                 =0.1cm,
    innerleftmargin             =0.1cm,
    innerrightmargin            =0.1cm,
    innertopmargin              =0.1cm,
    innerbottommargin           =0.1cm
}

\newcommand{\pankti}{\textsc{pankti}\xspace}
\newcommand{\reviseadd}[1]{\textcolor{black}{#1}}

\title{Production Monitoring to Improve Test Suites}
\author{Deepika Tiwari}
\author{Long Zhang}
\author{Martin Monperrus}
\author{Benoit Baudry}
\affil{KTH Royal Institute of Technology, Sweden}

\date{September 2020}

\begin{document}

\maketitle

\begin{abstract}
In this paper, we propose to use production executions to improve the quality of testing for certain methods of interest for developers. These methods can be methods that are not covered by the existing test suite, or methods that are poorly tested.
We devise an approach called \pankti which monitors applications as they execute in production, and then automatically generates differential unit tests, as well as derived oracles, from the collected data. \pankti's monitoring and generation focuses on one single programming language, Java. We evaluate it on three real-world, open-source projects: a videoconferencing system, a PDF manipulation library, and an e-commerce application. We show that \pankti is able to generate differential unit tests by monitoring target methods in production, and that the generated tests improve the quality of the test suite of the application under consideration.
\end{abstract}

\section{Introduction}

\IEEEPARstart{S}{oftware} developers write unit tests to assess programs with respect to an expected behavior captured in the form of a test oracle. 
Yet, the development of strong test suites is challenging: the selection of test data from complex input spaces is hard \cite{chen2008upper}, the specification of good test oracles is costly \cite{barr2014oracle,CatolinoPZF19}, and the assessment of test quality is time-consuming~\cite{Kochar17,Grano20}.
Test improvement has recently emerged as one way to assist developers in these tasks \cite{danglot2019snowballing}. The key idea is to combine developer-written tests with automatic analysis and generation techniques to complement these developer-written tests.
For example, one can consolidate test oracles with additional method calls \cite{xie2006augmenting}, or one can augment test inputs through operational abstraction \cite{harder2003improving}.
Our intuition is that we can also  use production data to cope with the weaknesses of developer-written tests.

In this paper, we introduce a novel approach called \pankti for improving test suites with observations collected in production. \pankti monitors an application in production, to collect execution data before and after certain methods of interest are executed, called target methods in this paper. \pankti subsequently extracts test input data and derived oracles \cite{barr2014oracle} from the collected data, and finally generates new test cases. 
The motivation of using production data is that production workloads may exercise behaviors not observed during the test execution \cite{wang2017behavioral}.
\pankti targets new test cases towards specific methods that are selected by the developers, according to what they need. For example, developers can decide to target methods  that are weakly-tested, as determined by  code coverage \cite{gligoric2013comparing,Hilton0M18}
or mutation testing \cite{petrovic2018state,MaOK06}.

\pankti works as follows, it first instruments the application to monitor the execution of target methods. The application is then deployed in production and executed according to some workload. For each invocation of a target method, the receiving object, the objects passed as arguments, as well as the returned object are serialized and stored. Once production monitoring data has been collected and persisted, \pankti deserializes the collected production objects to extract test inputs and synthesize derived oracles \cite{barr2014oracle} that capture the behavior observed in production. The last step consists in assembling the input and the oracle in a new differential unit test \cite{elbaum2006carving} for the target method, that is, a test that is able to capture behavioral change with respect to a reference version. The essence of \pankti is to recreate, in the generated differential unit test, the behavior observed in production. The final step in the \pankti test generation process consists in running the new tests to determine if indeed the overall test quality improves.

We implement \pankti for Java and evaluate it on three real-world, open-source, complex software systems: a videoconferencing system called Jitsi, a PDF manipulation library called PDFBox, and an e-commerce application called Broadleaf. These projects have between 28K and 729K lines of code and at least one thousand stars on GitHub. For evaluation purposes, we select as targets the methods that are found weakly-tested with respect to extreme mutation analysis \cite{niedermayr2016will}. 
For these experiments, \pankti targets a total of $86$ weakly-tested methods, and observes $122,194$ invocations of these methods in production. With the objects collected from these invocations, \pankti generates $14,222$ differential unit tests, of which $13,878$ ($97.6\%$) tests pass. Thanks to these \pankti-generated tests, the test quality of $53$ of $86$ ($61.6\%$) target methods is found to have improved per the considered test adequacy criterion. These results show that \pankti is able to automatically transform data observed in production into differential unit tests  that improve the quality of a test suite.
To sum up, our contributions are:
\begin{itemize}
\item \pankti, a  tool that monitors programs in production and uses the observed data to automatically generate differential unit tests that target weakly-tested methods;
\item An evaluation of the tool on three notable, real-world, open-source Java projects run in production: a videoconferencing system, a PDF manipulation library, and an e-commerce system;
\item A publicly available open-source implementation for Java\footnote{\url{https://github.com/castor-software/pankti}}, and open science experimental data\footnote{\url{https://doi.org/10.5281/zenodo.4298604}} for reproducibility and extension by further research.
\end{itemize}

In \autoref{sec:approach}, we describe our approach for test generation by monitoring programs in production. \autoref{sec:evaluation} describes our evaluation methodology, \autoref{sec:results} describes experimental results, \autoref{sec:discussion} discusses key insights and challenges from these results, \autoref{sec:related-work} discusses related work, and \autoref{sec:conclusion} concludes the paper. 

\section{The Pankti Approach to Test Generation}\label{sec:approach}

This section describes the details of our technical contribution. Our tool, called \pankti, exploits observations in production in order to generate new tests that complement the test suite of a Java application.


\subsection{Overview}

\begin{figure*}
\centering
\captionsetup{justification=centering}
\includegraphics[scale=0.57]{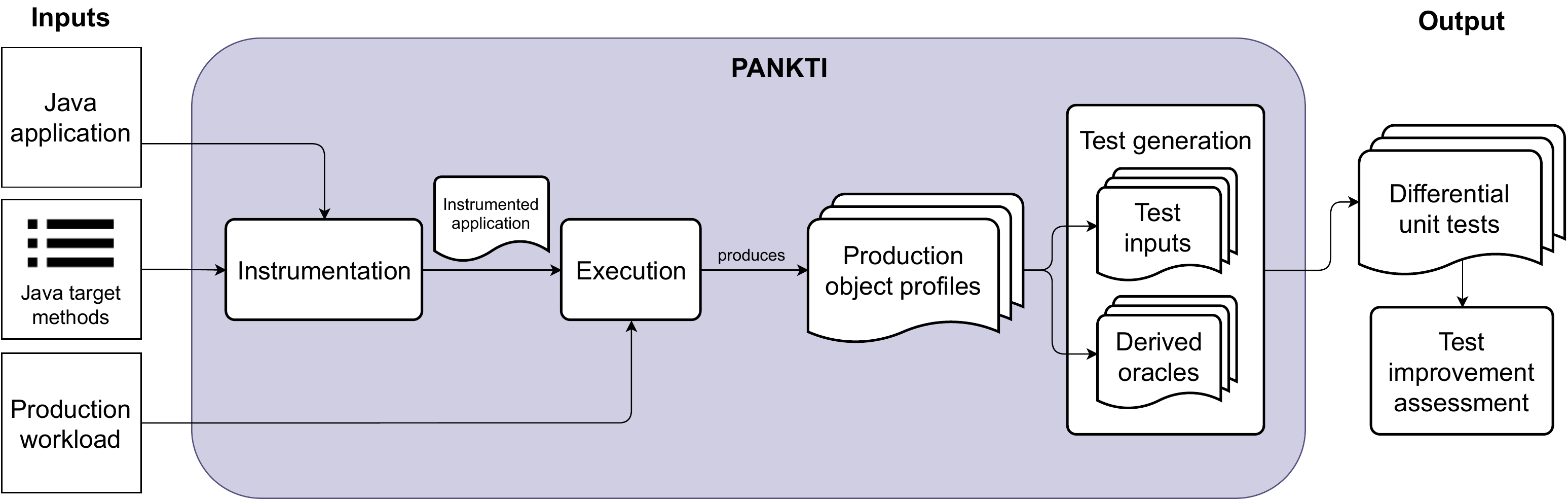}
\caption{The \pankti pipeline: Transforming production workloads into differential unit tests to improve testing of target methods according to a test adequacy criterion.}
\label{fig:pipeline}
\end{figure*}

The key concepts to describe \pankti are defined below.

\textbf{Test improvement:}
Test improvement is the process of strengthening an existing manually-written test suite, to enhance a specific, measurable test adequacy criterion \cite{inozemtseva2014coverage, zhang2015assertions}. This is the main goal of \pankti.

\textbf{Target methods:}
\pankti focuses on specific methods, for which it will generate new tests. Developers are responsible for selecting these methods based on their test adequacy criterion of choice. For example, \pankti can target methods in the code that are not fully covered by the test suite. We discuss target method selection in \autoref{sec:target-selection}.

\textbf{Production workload:}
In our study, a production workload is a sequence of inputs and interactions given to the program in production over a certain period \cite{tripleagent}. For instance, a production workload of a PDF reader is to open and scroll over a PDF file.

\textbf{Differential unit test:}
\pankti is designed to generate unit tests that are small, clear, and capture a targeted test intention.
The \pankti unit tests all include an oracle that is based on the behavior observed in production. This production-based oracle can be considered as a
derived test oracle per the classification of Barr \textit{et al.} \cite{barr2014oracle}. 
These unit tests aim to detect behavioral changes between the version monitored for test generation and upcoming revisions of the program, and are thus called ``differential unit tests'' by Elbaum \textit{et al.} \cite{elbaum2006carving}. To sum up,  \pankti generates differential unit tests with a derived oracle based on production data.

\textbf{Production object profile:}
When \pankti monitors the execution of an application under consideration with a specific workload, it collects one production object profile each time one of the target methods is invoked. We define an object profile as a quadruplet $<method, receiving, parameters, result>$ as follows:
given a target $method$ invocation, we monitor the $receiving$ object that the method is invoked on, with the objects or values passed as $parameters$ to the method, if any, as well as the $result$ object or value  returned from the method. 

\autoref{fig:pipeline} gives an overview of the \pankti pipeline for test generation. Our approach operates over three inputs: the source code of an application; a set of methods for which developers wish to get new tests; and one production workload. The output of \pankti is a collection of differential unit tests for the target methods that have been invoked at least once while executing the application.

The workflow of \pankti is structured in three phases. First, \pankti instruments each target method with binary instrumentation, to collect runtime data about its execution context. Second, \pankti executes the application with the input workload and collects runtime object profiles for each invocation of the target methods.
Third, \pankti generates differential unit tests for the target methods. When the target methods are selected based on a test adequacy criterion, \pankti selects the generated tests that improve the test quality for the method, according to that criterion.
The key concepts of \pankti are illustrated in the next section, and the different steps of our pipeline are presented in detail in the following sections.

\subsection{Selection of Target Methods}\label{sec:target-selection}

\pankti performs generation of differential unit tests for a subset of the methods of the application, this subset is chosen by the developer and passed as input configuration to \pankti. 
Focusing the generation on specific methods is important to ensure the relevance of the generated tests for the developers.

There are different ways for developers to choose a set of target methods. For example, developers might want to target the methods that implement the latest released feature, or to target the methods added or modified in the most recent commit  \cite{danglot2020approach}. Developers may manually specify the targets based on any combination of factors \cite{Kochar17}.
If there exists a test suite for the application, developers can pick as targets the methods that are not covered by this test suite \cite{wang2017behavioral}, or choose targets based on test inadequacies \cite{chen2020revisiting}. For example, a method may be considered as a target if it is covered by the test suite, but has surviving mutants \cite{petrovic2018state}, or if no test fails if its body is removed \cite{vera2019comprehensive}.
The test generation process of \pankti does not depend on how the target methods are selected.

\subsection{Working Example}\label{sec:working-example}
Here we illustrate the concepts of \pankti with a concrete example, taken from PDFBox\footnote{\url{https://github.com/apache/pdfbox/tree/2.0.21}}, an open-source application to work with PDF files \cite{butler2020maintaining}.
\autoref{lst:pt-method} shows a method called \texttt{getName} extracted from PDFBox. This method accepts four integer arguments, which are used as identifiers for the font name, platform, encoding, and language, respectively. It returns the name of the font corresponding to the integer identifiers, or \texttt{null} if the identifiers are not found.
We say that \texttt{getName} is inadequately tested by the original test suite of PDFBox, since replacing its body with the statements \texttt{return null;}, \texttt{return "A";}, or \texttt{return "";} is undetected by the $7$ existing test cases that trigger its execution \cite{vera2019comprehensive}. This means that \texttt{getName} is weakly-tested because, in spite of being covered by multiple tests, its output is not specified by a single oracle in the test suite \cite{5770598}.

\begin{figure}
\centering
\includegraphics[width=\columnwidth]{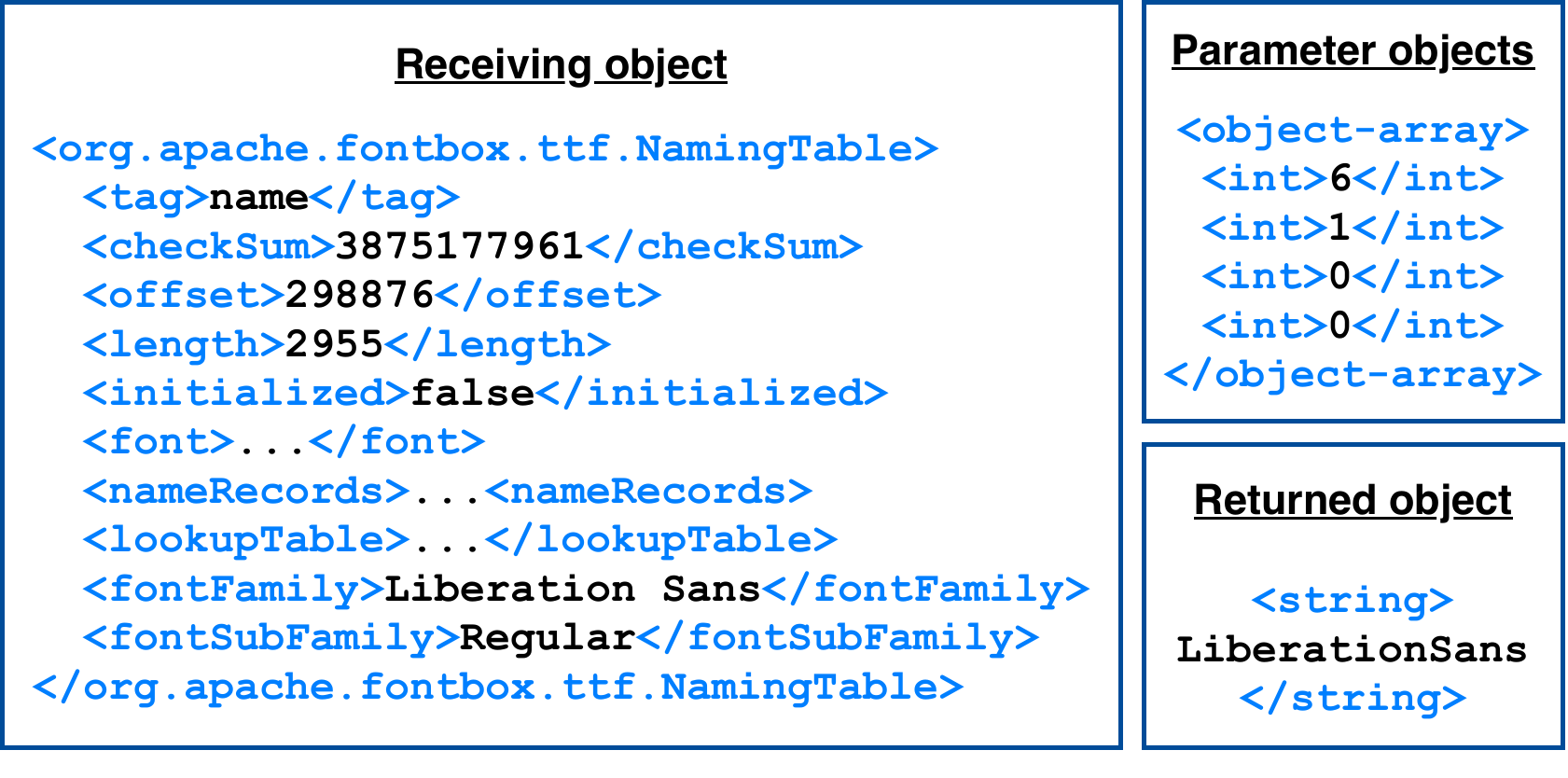}
\caption{A production object profile for the target method in \autoref{lst:pt-method}}
\label{fig:production-object-profile}
\centering
\vspace{-10pt}
\end{figure}

When we run PDFBox with a production workload, we observe that \texttt{getName} gets invoked $650$ times, and we collect $563$ production object profiles.
One of these object profiles is illustrated in \autoref{fig:production-object-profile}. It includes the $receiving$ object, which is the \texttt{NamingTable} object that \texttt{getName} is invoked on, the four integers passed as $parameters$ to it, and the \texttt{String} object returned as a $result$ from this invocation of \texttt{getName}. All the elements of the object profile are serialized in XML format.

\begin{lstlisting}[language=Java, belowskip={-10pt}, label={lst:pt-method}, caption={A target method for \pankti}, float]
String getName(int nId, int pId, int eId, int lId) {
  Map<Integer,Map<Integer,Map<Integer,String>>> platforms = lookupTable.get(nId);
  if (platforms == null) return null;
  
  Map<Integer,Map<Integer,String>> encodings = platforms.get(pId);
  if (encodings == null) return null;
  
  Map<Integer,String> languages = encodings.get(eId);
  if (languages == null) return null;
  
  return languages.get(lId);
}
\end{lstlisting}

\autoref{lst:generated-method} shows a differential unit test generated for \texttt{getName}, based on the production object profile illustrated in \autoref{fig:production-object-profile}. In the generated test, each element in the object profile is deserialized into its corresponding object. The generated assertion checks the equality between the \texttt{String} object as observed in production (expected value for this differential unit test), and the result returned from \texttt{getName} invoked on the deserialized \texttt{NamingTable} object, passing the same integer parameters.

\subsection{Execution with Observability}

A unique feature of \pankti consists in observing the execution of a program in production to collect data for test generation. This execution with observability is articulated around the following three steps.

\subsubsection{Code Instrumentation}
\label{sec:instrumentation}

The target methods under consideration are instrumented.
The instrumentation of target methods consists of adding  all the bytecode instructions required to monitor the execution context, before and just after the method invocation. For instance, the bytecode of the method \texttt{getName} in \autoref{lst:pt-method} is transformed into \autoref{lst:instr-method}, post-instrumentation. The highlighted lines in the excerpt in \autoref{lst:instr-method} show the instructions added to the original bytecode of \texttt{getName}.
The probe before the method invocation collects a snapshot of the state of the receiving object, i.e., the object the method is being invoked on, as well as the state of any object passed as parameter to the method. The probe after the invocation, saves the object returned by the method.

\begin{lstlisting}[style=bytecode, numbers=none, belowskip=-10pt, label={lst:instr-method}, caption={Excerpt showing bytecode transformation of the target method in \autoref{lst:pt-method}}, float]
public java.lang.String getName(int, int, int, int);
  Code:
  ...
  @197: invokestatic  #67@ // Method
         se/kth/castor/pankti/instrument/plugins/
         NamingTableAspect$TargetMethodAdvice.onBefore
  202: aload_0
  203: getfield       #24 // Field lookupTable:Ljava/util/Map;
  ...
  @306: invokestatic  #96@ // Method
         se/kth/castor/pankti/instrument/plugins/
         NamingTableAspect$TargetMethodAdvice.onReturn
  309: areturn
\end{lstlisting}

\pankti depends on the instrumentation of target methods at the bytecode level in a Java application in production. The key challenge here is to find a robust and reliable framework that can perform this instrumentation at scale, in the production environment.

\subsubsection{Collection of Production Object Profiles}
When the application executes under a specific workload, the instrumentation code for each target method, such as the one in \autoref{lst:instr-method}, generates one production object profile  $<method, receiving, parameters, result>$, each time the target method is invoked. \pankti collects production object profiles in real time, and saves them to the disk.
Production objects can be large and complex, with several levels of nesting with sub-objects. Serialization of very deeply nested objects is known to be hard \cite{java_serialization,6215346}. For this reason, \pankti requires state-of-the-art serialization technology, which will be presented in \ref{sec:implementation}. Another crucial requirement of \pankti is to save the objects in a human-readable format, because they are meant to be used in generated tests read by developers.

\subsubsection{Selection of Object Profiles}\label{sec:selection-object-profiles}
During a pilot experiment, we made two important observations: some target methods are invoked thousands of times as the application executes; some of the serialized objects are very large.
For instance, the method \texttt{toUnicode(int)} in PDFBox was invoked $30,840$ times with our selected production workload, and the \texttt{PDTrueTypeFont} object it was invoked on had nesting up to $13$ levels deep. In production, there is a limited amount of space available to save object profiles.
Based on these observations, \pankti uses a threshold on the maximum size of serialized object profiles. 
The size threshold is one configuration parameter of \pankti, for which the default value is $200$ MB. This implies that object profiles are collected for each invocation of a target method, provided this threshold is not encountered for that method.

\subsection{Generation of Test Cases}
\label{sec:test-gen}

The test generation phase of \pankti processes the object profiles collected in production to synthesize differential unit tests for the target methods that have been invoked at least once.

We have observed that a target method is often invoked in the same context and the set of collected production object profiles for one specific target method contains redundant data. For instance, the method \texttt{getName} in \autoref{lst:pt-method} is invoked $650$ times during execution with our selected workload. We collect production object profiles for $563$ invocations, as per the size threshold discussed in the previous section. Of these $563$ object profiles, $273$ are unique.
The first step for test generation is to select unique profiles.

Second, for each unique profile, \pankti generates a differential unit test that follows a systematic template. 
First, the test case deserializes each part of the object profile: the receiver, the parameters, and the result object. 
Second, the test case invokes the $method$ on an object initialized with the deserialized $receiving$ and passes the set of deserialized $parameters$ to the method call. 
Third, we generate an assertion that expects the method invocation to return the same $result$ as observed in production. The serialized representations of objects can vary in size, depending on the complexity of the corresponding object. For the sake of readability, \pankti includes smaller representations directly in the generated test, but generates resource files for serialized representations that are more sizable. The corresponding objects are then deserialized from these files during the execution of the generated test. This is also the methodology adopted by \cite{kvrikava2018tests}.
For example, \pankti generated the test case shown in \autoref{lst:generated-method}, for the method in \autoref{lst:pt-method}, based on the  production object profile illustrated in  \autoref{fig:production-object-profile}. Lines $4$ and $5$ in the test fetch the $receiving$ object that was observed in production from a resource file and deserialize it. Lines $6$ and $7$ deserialize the $result$. Line $8$ is the test oracle that asserts an equality between the $result$ that was observed in production and the value returned by the target method \texttt{getName} when it is invoked with the $parameters$ as obtained from the production object profile.

Deserialization of arbitrary objects is a hard problem \cite{java_serialization,6215346}, and can sometimes result in an object that is not identical to the one observed in production. Since \pankti fully relies on a full serialization-deserialization cycle, we need production grade serialization technology.

\subsection{Test Improvement Assessment}\label{sec:improvement-assessment}
The final step of the \pankti pipeline assesses and filters the test cases generated for each target method. 
Here, we check three aspects:
1) that the generated test passes (see discussion in \autoref{sec:failing-tests}),
2) that it is not flaky\footnote{we run it $5$ times per \cite{8453104}}, and
3) that the test improves the test adequacy criterion originally used to select the target methods.

The latter action is performed only if the application has a test suite and if the target methods were selected based on a test inadequacy criterion. In this case, the quality of the test suite before and after the test generation process is compared, per the test inadequacy criterion. If the quality is found to have improved, \pankti is considered successful in generating a useful test.

\begin{lstlisting}[label={lst:generated-method}, belowskip=-10pt, caption={A generated differential unit test for the target method in \autoref{lst:pt-method}}, float]
@Test
public void testGetName() {
  // generated test for target NamingTable.getName 
  File fileReceiving = new File("receiving.xml");
  NamingTable receivingObject = deserializeObject(fileReceiving);
  
  String returnedObject = "<string>LiberationSans</string>";
  String expectedObject = deserializeObject(returnedObject);
  
  assertEquals(expectedObject, 
                   receivingObject.getName(6, 1, 0, 0));
}
\end{lstlisting}

\subsection{Implementation}\label{sec:implementation}

\pankti is implemented in Java.
By default, \pankti focuses on generating tests for public and non-static methods. The focus on public methods provides a clear intention for the generated tests, similar to what developers do in the original test suite. Since \pankti collects object profiles to generate the test inputs and the oracle, we do not target static methods in order to focus the test generation on the invocation of methods on objects and not classes.

\pankti uses Glowroot\footnote{\url{https://glowroot.org/}} for code instrumentation and observability. Glowroot is an open-source, robust, and lightweight application performance management tool. \pankti extends its production monitoring and instrumentation capabilities using its Plugin API. 

The serialization and deserialization of production object profiles rely on the XStream library\footnote{\url{http://x-stream.github.io/}}. XStream is a state-of-the art library for serialization, is very robust, can handle arbitrarily complex objects, and allows the registration of custom converters. This library has also been proven to be an effective serialization tool in previous research \cite{elbaum2006carving,pasternak2009genutest,artzi2008recrash}. Through a parameter, \pankti can be configured to generate tests with serialized object strings in XML, JSON, and all other formats supported by XStream.

The test synthesis phase relies on Spoon \cite{pawlak:hal-01078532} for test code generation. Spoon is a code analysis and transformation library for Java programs. \pankti leverages the code generation features of Spoon to generate tests that are syntactically correct.

In the last phase, we assess the quality of the tests by compiling and running them. If the target methods are selected based on a test adequacy criterion, we determine if the \pankti tests improve the test suite with respect to the same criterion.

\pankti is made publicly available for facilitating reproduction and future research on the topic, at \url{https://github.com/castor-software/pankti/}.

\section{Evaluation Methodology}
\label{sec:evaluation}

This section describes our approach to evaluate \pankti on real-world software systems. We discuss the criteria used to select the study subjects, their target methods and workloads, as well as the metrics computed for these projects.

\subsection{Study Subjects}
\label{sec:study-subjects}
We perform a case-based research study \cite{flyvbjerg2006five} to evaluate \pankti. 
We select a set of study subjects according to the following criteria:
1) The project is a real-world, open-source product implemented in Java, built with Maven, and has a test suite;
2) The project has more than $1000$ stars and more than $1000$ commits, which is indicative of a popular and mature project;
3) The project builds, the test suite passes, and the project can be deployed with the computing resources of our research lab;
4) A realistic production workload for the project can be executed in our laboratory environment.

We manually search for projects on GitHub that meet all of the above criteria, and select $3$ of them.
Jitsi/Jicofo is a component used in the Jitsi Meet videoconference system, a high-quality conferencing solution which supports audio, video, and text communication. Jicofo is responsible for maintaining the connection of the participants of a Jitsi Meet conference.
PDFBox is a robust and mature project by the Apache Software Foundation for producing and reading PDF documents such as legal decisions, invoices, and contracts. This PDF application provides capabilities such as digital signatures, image conversion, etc. It includes standalone command-line tools that can be invoked on PDF and text documents \cite{butler2020maintaining}.
Our third subject is BroadleafCommerce (henceforth referred to as Broadleaf), which is an enterprise e-commerce framework based on Spring. Broadleaf provides all necessary components for a web shopping experience, such as user and product management, a front-end website, and API services.

The version, number of stars (\#STARS), number of commits (\#COMMITS), and lines of code (LOC) for each of these projects are shown in \autoref{tab:selected-projects}. The three projects have between 28K and 729K lines of code.

\begin{table}
\newcounter{mpFootnoteValueSaver}
\setcounter{mpFootnoteValueSaver}{\value{footnote}}
\centering
\caption{Real-world projects used in our experiments}
\label{tab:selected-projects}
\begin{tabularx}{\columnwidth}{llRRRR}
\toprule
\textbf{PROJECT}& \textbf{VERSION}& \textbf{\#STARS}& \textbf{\#COMMITS}& \textbf{LOC}\\
\midrule
Jitsi/Jicofo& stable-4857\footnotemark& 14K& 1,248& 28.32K\\
PDFBox& 2.0.21\footnotemark& 1.3K& 8,244& 728.8K\\
Broadleaf& 6.1.4-GA\footnotemark& 1.4K& 16,964& 197.5K\\
\bottomrule
\end{tabularx}
\end{table}
\stepcounter{mpFootnoteValueSaver}
\footnotetext[\value{mpFootnoteValueSaver}]{\url{https://github.com/jitsi/jicofo/tree/stable/jitsi-meet_4857}}
\stepcounter{mpFootnoteValueSaver}
\footnotetext[\value{mpFootnoteValueSaver}]{\url{https://github.com/apache/pdfbox/tree/2.0.21}}
\stepcounter{mpFootnoteValueSaver}
\footnotetext[\value{mpFootnoteValueSaver}]{\url{https://github.com/BroadleafCommerce/BroadleafCommerce/tree/broadleaf-6.1.4-GA}}

\subsection{Experiments with Production Workloads}
\label{sec:workloads}

To generate differential unit tests using production data for each of our study subjects, we execute them in production conditions that are realistic, relevant, and reliable. This section describes our choice of workloads for each of the study subjects.

\subsubsection{Production workloads for Jitsi/Jicofo}

We hold a videoconference meeting to obtain a Jitsi/Jicofo workload. For this, we deploy the full Jitsi stack using the \texttt{docker-jitsi-meet} project. We create a meeting room for three members of the research lab for the experiment. The members join this meeting room from different IP addresses to have an hour-long meeting. During the meeting, all the participants turn on their camera and microphone. Each participant also sends at least one text message via the chat box.

\subsubsection{Production workloads for PDFBox}
\label{sec:prod-workload-pdfbox}
As documentation and archival become more digitized across all domains, the Portable Document Format, or PDF, has emerged as a standard to ensure platform-independence and interoperability with multiple user environments. To select realistic and non-trivial workloads to experiment with PDFBox, we shortlist five PDF documents from GovDocs1 \cite{GARFINKEL2009S2}, an online corpora of PDF files made available for research and analysis. The same protocol has also been followed in previous research \cite{kuchta2018correctness}. In order to have a good trade-off between the diversity of the files and the execution time for the experiments, we select the PDF documents based on the following criteria: 1) They have between 1 to 5 pages of text;  2) They include at least one image or photograph; and 3) Their size does not exceed $1.5$ MB. 
The five documents that we use as part of the PDFBox workload are available as part of our replication package.

For each of the five selected documents, we invoke ten PDFBox features, which include the encryption of the PDF document with a password, decryption of an encrypted PDF document, extraction of text and images from a PDF document, etc. In summary, the workload to evaluate \pankti on PDFBox is composed of $5 \times 10$ different invocations to the PDFBox command line tool.

\subsubsection{Production workloads for Broadleaf}

The workloads for Broadleaf consist of general steps to purchase products online as an end-user: a user visits the homepage of the website first, then registers a new account and logs in, then views the web page to add some products into the shopping cart, and finally the user checks out the products and logs out.
There are $871$ HTTP requests in the workload, comprising of $865$ GET requests and $6$ POST requests. The GET requests are used to fetch product information and necessary files, such as JavaScript and CSS files. The POST requests are mainly for user registration, logging in, checking out products, and logging out.

\subsection{Experiments with System Tests}\label{sec:system-tests}
System tests are typically designed to verify end-to-end uses of applications. They may exercise multiple and complex scenarios \cite{messaoudi2021log} or use cases  \cite{wang2015automatic} through the system.
It is possible to generate differential unit tests from system tests, as proposed by the seminal papers of Elbaum \textit{et al.} \cite{elbaum2006carving} and Saff \textit{et al.} \cite{saff2005automatic}.
Inspired by this related work in the area of test improvement, we perform an experiment where we use \pankti with system tests, instead of a production workload.

The goal of this experiment is to demonstrate the ability of \pankti to generate unit tests by observing the execution of systems tests. So first, we need to select system tests. However, there is no strict definition of what constitutes a system test. Per previous research, we note that
system tests and unit tests differ in the extent of the application they exercise, or the number of lines of code they cover \cite{4639304}.  
Consequently, as actionable definition, we define system tests as those tests in the original test suite of our study subjects that cover at least $1500$ lines of code in the application (this threshold of $1500$ lines of code to identify system tests was used in previous research \cite{danglot2019automatic}). We measure the line coverage of each test with JaCoCo\footnote{\url{https://www.eclemma.org/jacoco/}}, in order to select the ones we consider as system tests.

We pass those system tests to \pankti, which collects object profiles whenever target methods are invoked during their execution. The collected object profiles from system tests are then used to generate differential unit tests.
The results of using \pankti to carve differential unit tests from system tests will be discussed in \autoref{sec:results-system-test}

\subsection{Target Methods for The Evaluation}
\label{sec:targets-evaluation}

For our evaluation, we use ``pseudo-tested methods'' as the test adequacy criterion \cite{niedermayr2016will,vera2019comprehensive} to select target methods in the three projects. A method is said to be pseudo-tested if it is invoked at least once when running the test suite, but no test fails if its body is removed. This means that no existing test is able to detect extreme transformations of these methods. Pseudo-testedness is a state-of-the-art criterion that developers care about, as evaluated in user studies \cite{vera2019comprehensive}. We use the Descartes\footnote{\url{https://github.com/STAMP-project/pitest-descartes}} tool to automatically identify pseudo-tested methods in the three study subjects.

This choice of targets for the evaluation implies a very clear way to assess the improvement provided by the \pankti differential unit tests. 
A test for a specific target method is considered to be an improvement if the target method is no longer pseudo-tested.

\subsection{Measurements}\label{sec:measurements}
In this section, we introduce the metrics that we collect as part of our case-based study and assessment of \pankti. 

\emph{Testing metrics. }First, we compute the following quantitative metrics capturing the test quality of the subjects:\\ 
\textbf{1) Test suite method coverage, TMCov}, represents the percentage of methods in the project that are covered by at least one test case in the original test suite.\\
\textbf{2) Test suite line coverage, TLCov}, represents the percentage of the lines of code in the project covered by at least one test case in the original test suite.\\
\textbf{3) Workload method coverage, WMCov}, determines the percentage of methods in the project that are invoked when we execute the project with our selected production workload.\\
\textbf{4) Workload line coverage, WLCov}, determines the percentage of lines of code in the project that are executed with the selected workload.\\
\textbf{5) Target methods, \#TARGET}, is the number of target methods in the project. In the case of our evaluation, this is the number of pseudo-tested methods, i.e., the methods that are covered by at least one test case in the original test suite, but no test case fails if their body is replaced with a statement returning a default value; and\\
\textbf{6) Effective target methods, \#EFF\_TARGET}, is the number of target methods for which we use \pankti to generate units tests. In the case of the experiments with production workloads, this is the number of pseudo-tested methods that are invoked when running the workload. In the case of the experiments with system tests, this is the number of pseudo-tested methods that are invoked when running the system tests.

Metrics TMCov and TLCov are indicators of the quality of the test suite of the project. Metrics WMCov and WLCov indicate the proportion of the project that is covered by our selected production workload. Metrics \#TARGET and \#EFF\_TARGET are computed specifically as starting points for \pankti. 
We use JaCoCo to obtain the coverage. 
\#TARGET is computed with Descartes \cite{vera2019comprehensive}, while \#EFF\_TARGET is the subset of \#TARGET covered during the experiments with production workloads and with system tests.

The second set of metrics we collect aim at assessing the ability of \pankti at transforming production data into differential unit tests.\\
\textbf{7) Target method invocations, \#INVOCATIONS}, is the number of invocations of an effective target method as the application executes with the production workload.\\
\textbf{8) Collected object profiles, \#COLLECTED}, is the number of production object profiles for each effective target method that are captured as the application executes with the selected workload.\\
\textbf{9) Unique object profiles, \#UNIQUE}, is the size of the subset of collected object profiles that includes unique combinations of the constituent elements.\\
\textbf {10) Object profile size, SIZE} is the size on disk of collected production object profiles.\\
\textbf{11) Differential unit tests, \#PANKTI\_TESTS}, is the number of test cases generated by \pankti using the \#UNIQUE object profiles.\\
\textbf{12) Passing test cases, \#PASSING}, is the number of generated tests that pass when executed.\\
\textbf{13) Failing test cases, \#FAILING}, is the  number of generated tests that do not pass when executed.\\
\textbf{14) PANKTI\_STATUS} (pseudo-tested or well-tested) is the classification of an effective target method, as obtained by Descartes, after the addition of the generated test to the test suite of the project. After adding the \pankti-generated tests in the application test suite, a target method is either still pseudo-tested, or, is not pseudo-tested anymore. In the latter case, we give it the status of \emph{well-tested}.

Metrics \#INVOCATIONS, \#COLLECTED, \#UNIQUE, and SIZE are intermediate outputs of \pankti as it monitors the target methods in the project in the production environment. Metrics \#PANKTI\_TESTS, \#PASSING, \#FAILING, and PANKTI\_STATUS are the final outputs of \pankti for the project, and are indicators of its success in generating differential unit tests for the target methods in the project.

\section{Experimental Results}\label{sec:results}
This section discusses our findings when running \pankti on our three study subjects. First, we present general data about the quality of the subjects' test suites. The subsequent sections discuss each case study separately.

\subsection{Descriptive Statistics}

\begin{table*}
\centering
\caption{Original test suite method and line coverage (TMCov, TLCov), workload method and line coverage (WMCov, WLCov),  number of pseudo-tested methods (\#TARGET) and number of pseudo-tested methods invoked when running the workload (\#EFF\_TARGET) for each study subject}\label{tab:computed-metrics}
\begin{tabular}{@{}lrrrrrr@{}}
\toprule
\textbf{PROJECT} & \textbf{TMCov} & \textbf{TLCov} & \textbf{WMCov} & \textbf{WLCov} & \textbf{\#TARGET} & \textbf{\#EFF\_TARGET} \\ \midrule
Jitsi/Jicofo & \begin{tabular}[c]{@{}r@{}}49.4 \%\\ (667 / 1,350)\end{tabular} & \begin{tabular}[c]{@{}r@{}}46.7 \%\\ (3,537 / 7,571)\end{tabular} & \begin{tabular}[c]{@{}r@{}}48.9 \%\\ (660 / 1,350)\end{tabular} & \begin{tabular}[c]{@{}r@{}}46.2 \%\\ (3,500 / 7,571)\end{tabular} & 29 & 29 \\ \midrule
PDFBox & \begin{tabular}[c]{@{}r@{}}54.8 \%\\ (6,049 / 11,042)\end{tabular} & \begin{tabular}[c]{@{}r@{}}53.5 \%\\ (34,653 / 64,787)\end{tabular} & \begin{tabular}[c]{@{}r@{}}21.6 \%\\ (2,390 / 11,042)\end{tabular} & \begin{tabular}[c]{@{}r@{}}21.0 \%\\ (13,630 / 64,787)\end{tabular} & 138 & 46 \\ \midrule
Broadleaf & \begin{tabular}[c]{@{}r@{}} 23.9 \%\\ (1,478 / 6,173)\end{tabular} & \begin{tabular}[c]{@{}r@{}}23.7 \%\\ (5,846 / 24,667)\end{tabular} & \begin{tabular}[c]{@{}r@{}}23.1 \%\\ (1,424 / 6,173)\end{tabular} & \begin{tabular}[c]{@{}r@{}}19.2 \%\\ (4,735 / 24,667)\end{tabular} & 32 & 11 \\ \bottomrule
\end{tabular}
\end{table*}

\autoref{tab:computed-metrics} summarizes the key metrics about the original test suite and about the workload: TMCov, TLCov, WMCov, WLCov, \#TARGET, and \#EFF\_TARGET (see  \autoref{sec:measurements}). Let us take the case of Jitsi/Jicofo as an example. The original test suite of Jitsi/Jicofo covers $49.4\%$ of the methods, and $46.7\%$ of the lines in the project. Our selected production workload for Jitsi/Jicofo covers $48.9\%$ of the methods and $46.2\%$ of the lines in the project. We find that 29 public and non-static methods are pseudo-tested in Jitsi/Jicofo (\#TARGET). This means that they are covered by at least one test in the suite, but no test is able to detect extreme mutations in these methods. When we instrument these 29 methods and observe Jitsi in production, we find that all $29$ of them get invoked (\#EFF\_TARGET). 

Per \autoref{tab:computed-metrics}, the original test suite of all the three projects varies in coverage, with a highest line coverage of 53.5\%.
Moreover, the coverage achieved by the workload is lower than the test coverage for all of the three cases. This observation is consistent with the findings of Wang \textit{et al.} \cite{wang2017behavioral}.

Pseudo-tested methods are present in all projects, confirming the results of Vera-Perez \textit{et al.} \cite{vera2019comprehensive} about the prevalence of such methods. These methods, covered by the original test suite, but weakly-tested, provide a false sense of trust in the test suite, which we aim at mitigating with new test cases generated by \pankti.

On executing each project with its corresponding workload, some of these pseudo-tested methods are indeed invoked in production. This is indicated by the column "\#EFF\_TARGET" in \autoref{tab:computed-metrics}.
This is the baseline that \pankti aims to improve: we want to generate tests such that some of those pseudo-tested methods become well-tested: 29, 46, and 11 target methods for Jitsi/Jicofo, PDFBox, and Broadleaf, respectively.

\input{table-experiment-results}

\subsection{Case Study 1: Jitsi/Jicofo}\label{sec:results-jicofo}

\begin{figure}
\centering
\includegraphics[width=\columnwidth]{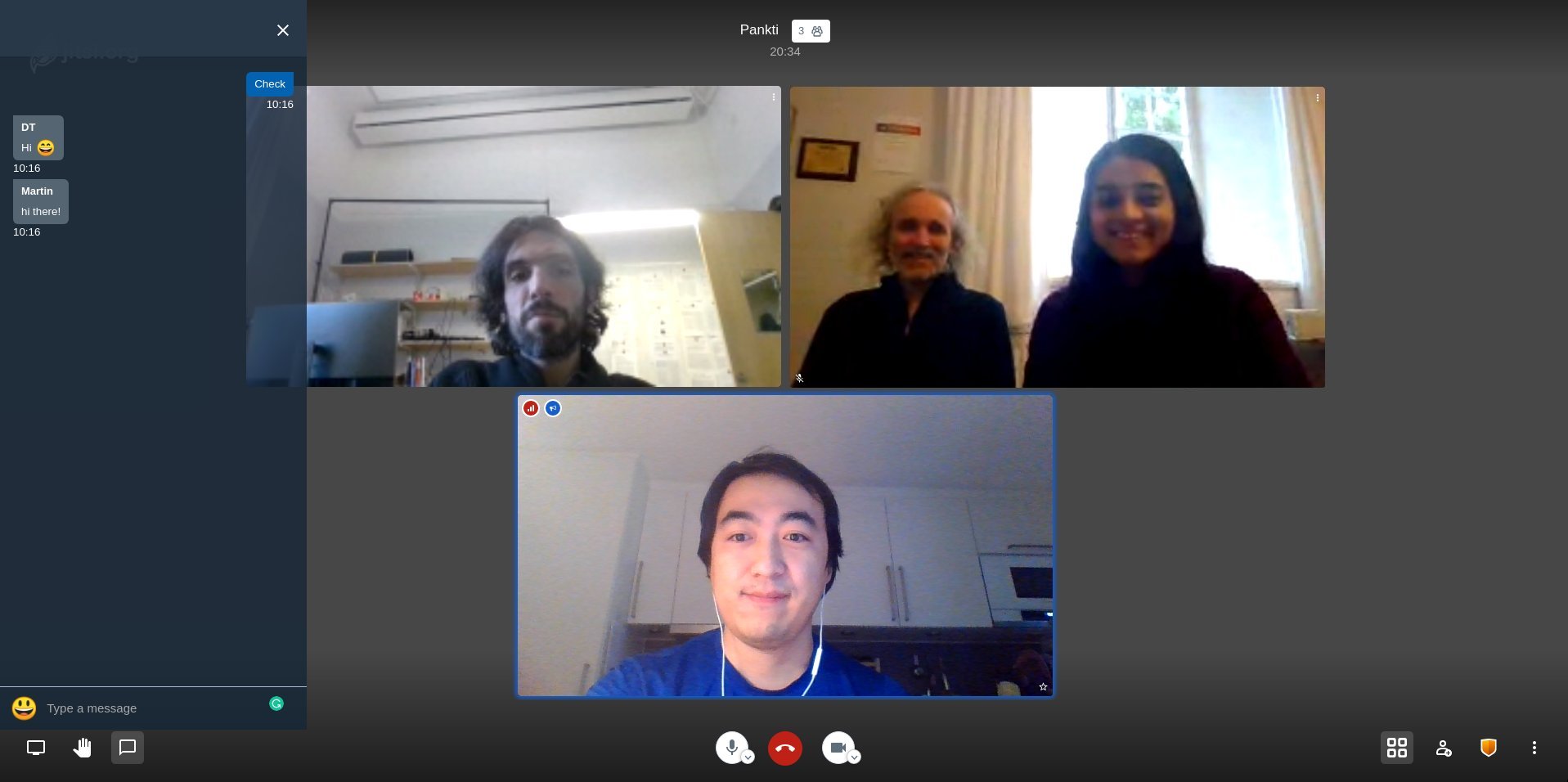}
\caption{A Jitsi Meeting with \pankti Attached}
\label{fig:jitsi-meeting}
\vspace{-10pt}
\end{figure}

\pankti targets 29 pseudo-tested methods in Jitsi/Jicofo for instrumentation. Then, we run Jitsi for a videoconference during which we collect object profiles. We, the authors, perform this videoconference to discuss this paper. \autoref{fig:jitsi-meeting} presents a screenshot of this meeting, where \pankti was attached to the Jitsi/Jicofo component. During the one-hour meeting, target methods keep getting invoked periodically, triggering object profile collection, yet no participant observes a degradation of Jitsi's user experience such as a delay, or a crash.

We now discuss the results of \pankti on Jitsi, presented in \autoref{tab:experiment-results}. \autoref{tab:experiment-results} is grouped by the study subject and sorted by the invocation count of the target methods. Each row corresponds to one effective target method, and includes the values of the metrics introduced in \autoref{sec:measurements}. For each of these methods, \autoref{tab:experiment-results} shows the number of times it was invoked with the production workload (\#INVOCATIONS), the number of object profiles corresponding to these invocations that are serialized (\#COLLECTED), the number of collected object profiles that are unique (\#UNIQUE), as well as the number of differential unit tests generated by \pankti (\#PANKTI\_TESTS). Of the tests generated, the number of tests that pass (\#PASSING), and the ones that do not (\#FAILING), are also indicated. The Descartes classification (PANKTI\_STATUS) of the target method after the addition of the generated tests to the test suite of the project is shown in the last column, it either remains the same as before, i.e., ``pseudo-tested'', or upgrades to ``well-tested''.

Rows 1 to 29 of \autoref{tab:experiment-results} show the experimental results for Jitsi. 
The 29 effective target methods are invoked 152 times during the experiment. \pankti collects 110 object profiles, 20 of which are unique. Based on the unique object profiles, \pankti generates 20 differential unit tests in total for Jitsi/Jicofo. All of the generated tests pass and improve the quality of the Jitsi/Jicofo test suite, transforming  $19/29$ ($65.5\%$) pseudo-tested methods into well-tested ones.

\begin{figure}
\centering
\includegraphics[width=\columnwidth]{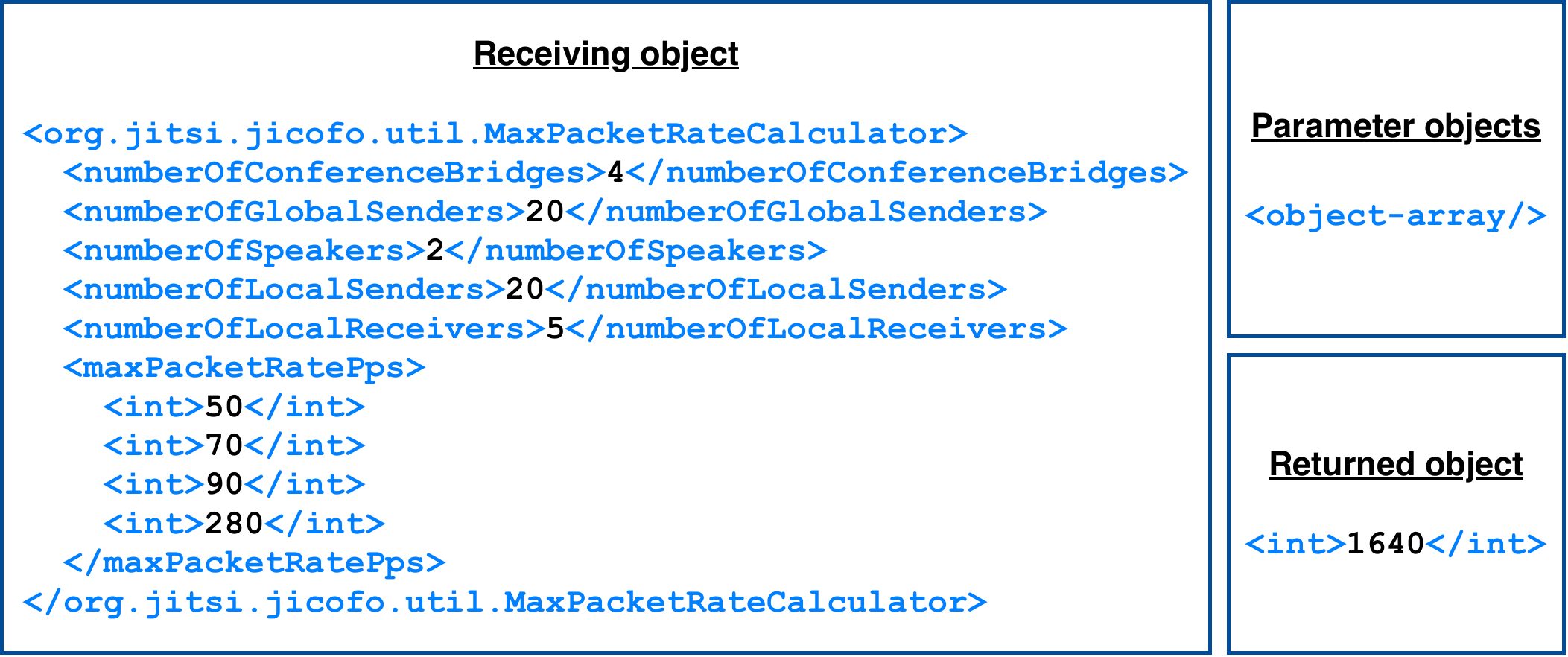}
\caption{A production object profile for the method \texttt{computeParticipantEgressPacketRatePps()} in Jitsi/Jicofo}
\label{fig:jicofo-object-profile}
\centering
\vspace{-15pt}
\end{figure}

Let us look at one successful case. \autoref{lst:pseudo-tested-method-jicofo} is the source code of method \texttt{computeParticipantEgressPacketRatePps} (row $27$ in \autoref{tab:experiment-results}) in class \texttt{MaxPacketRateCalculator}. The method is covered by developer-written unit tests, but is identified as pseudo-tested: if the method body is replaced with \texttt{return -1;}, \texttt{return 0;}, or \texttt{return 1;}, no test in the original test suite fails.
By observing the behavior of this method in production, \pankti collects the object profile shown in \autoref{fig:jicofo-object-profile}. A new differential unit test is then generated based on this profile, shown in \autoref{lst:test-case-jicofo}. The generated test includes an explicit assertion  for the behavior of \texttt{computeParticipantEgressPacketRatePps}. After adding this generated differential unit test to the test suite, the method is not pseudo-tested anymore.

\begin{lstlisting}[language=Java, belowskip=-10pt, label={lst:pseudo-tested-method-jicofo}, caption={A target method called \texttt{computeParticipantEgressPacketRatePps} in Jitsi/Jicofo}, float]
public int computeParticipantEgressPacketRatePps() {
  return (numberOfSpeakers * maxPacketRatePps[0]
          + (numberOfGlobalSenders - 2)
          * maxPacketRatePps[1] + maxPacketRatePps[3]);
}
\end{lstlisting}

\begin{lstlisting}[language=Java, belowskip=-10pt, label={lst:test-case-jicofo}, caption={A generated differential unit test for the \texttt{computeParticipantEgressPacketRatePps} method}, float]
@Test
public void testCPEgressPacketRatePps() {
  // content of the serialized receiving object from Fig. 4
  String receivingObjectStr = "<org.jitsi.jicofo.util.MaxPacketRateCalculator> ... </org.jitsi.jicofo.util.MaxPacketRateCalculator>"; 
  
  MaxPacketRateCalculator receivingObject = deserializeObject(receivingObjectStr);
  
  assertEquals(1640, receivingObject.computeParticipantEgressPacketRatePps());
}
\end{lstlisting}

During the object profile collection phase, \pankti collects $110$ object profiles for $20$ out of $29$ ($70.0\%$) methods. This corresponds to a total of $9.79$ KB collected data, for object profiles that vary between $0.48$ KB and $0.50$ KB, with a median value of $0.49$ KB. For the duration of our experiment, none of the methods reach the threshold of the file size ($200$ MB) for their object profile collection.

We observe that \pankti collects no object profile for 9 methods, and does not collect all the profiles for 3 methods (rows 1, 3, and 4).  This occurs when \pankti faces extreme situations for object serialization at runtime:
1) A \texttt{ConcurrentModificationException} happens if XStream tries to serialize an object that is being changed. This may happen if the object contains some fields which are accessed by multiple threads frequently.
2) A \texttt{``no converter specified''} exception happens if XStream does not know how to serialize some members in the object. This case only happens when XStream fails to serialize class \texttt{NetworkAddressManagerServiceImpl} in the package \texttt{net.java.sip.communicator.impl.netaddr} of Jitsi/Jicofo. This is due to the fact that this class contains fields that are related to system resources such as sockets and threads, which are fundamentally unserializable.
3) a \texttt{``security rule violation''} happens if XStream refuses to serialize a specific class which is protected by security code.
Overall, this clearly show that serialization of arbitrary objects is fundamentally hard, and not yet solved by the state-of-the-art of serialization. After the profile collection phase, \pankti successfully generates test cases for all of the covered methods. All these new test cases pass.

\begin{mdframed}[style=mpdframe,frametitle=Highlight from the Jitsi/Jicofo Experiment]
The experiment on Jitsi/Jicofo uses a production workload for videoconferencing, an essential feature of remote working and social interactions in 2020 and 2021. \pankti successfully collects $110$ object profiles for $20$ methods. \pankti generates $20$ new test cases that improve the testing of $19$ out of $29$ target methods of Jitsi. This experiment shows that \pankti does not perturb user experience, users can have a smooth videoconference while generating tests. \end{mdframed}

\subsection{Case Study 2: PDFBox}
\label{sec:results-pdfbox}
There are $138$ pseudo-tested methods in PDFBox. The production workload covers $46$ of these methods, which all become effective target methods for \pankti. These methods are included in rows 30 to 75 in \autoref{tab:experiment-results}. From the $121,175$ total invocations of these target methods in production, which occur throughout the course of our experiment, we collect $57,563$ production object profiles.

\begin{figure}[h]
\vspace{-10pt}
\centering
\includegraphics[width=\columnwidth]{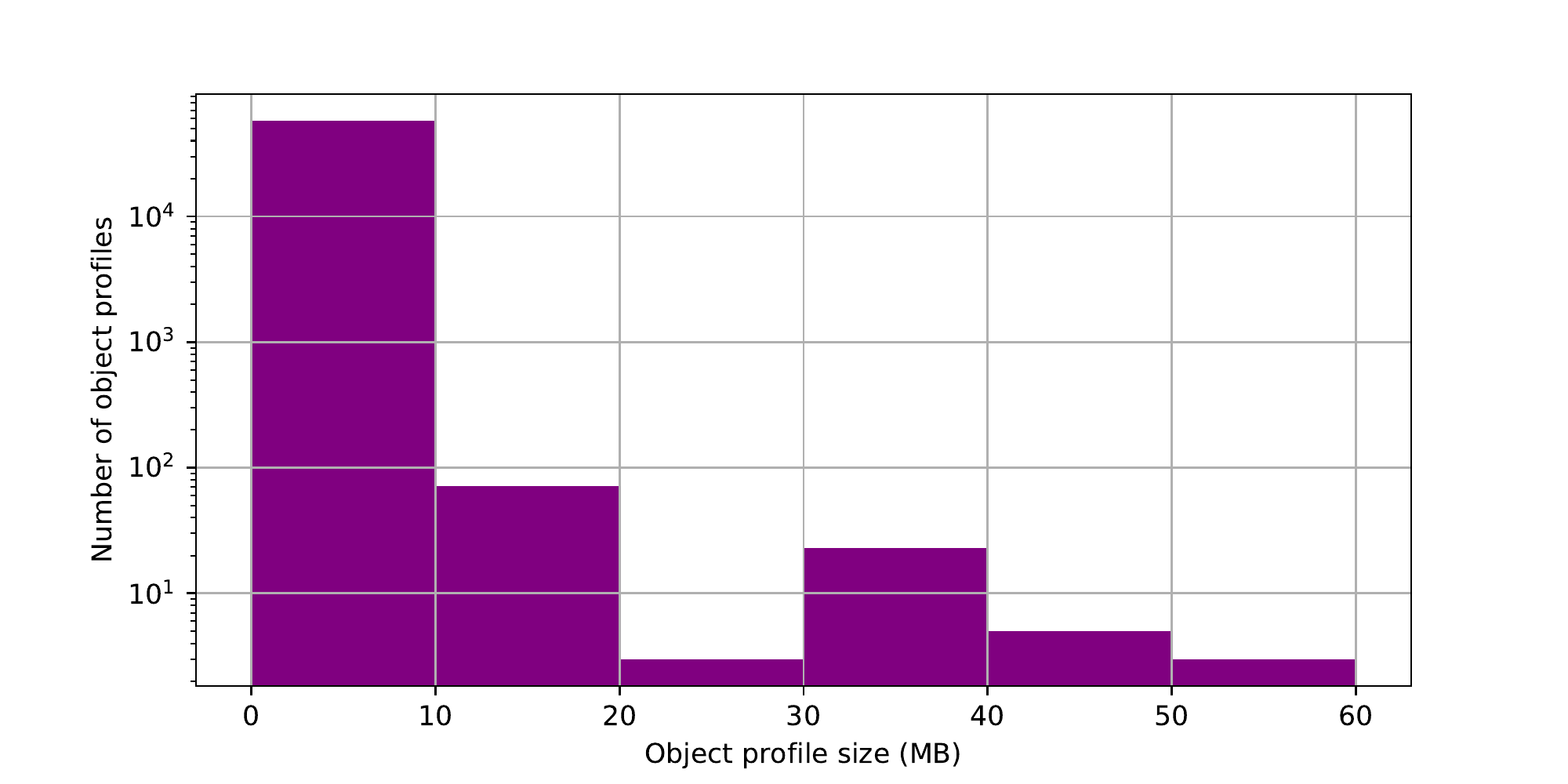}
\caption{The distribution of the collected object profile sizes for PDFBox over $57,563$ profiles from $46$ methods}
\label{fig:size-distribution}
\centering
\vspace{-10pt}
\end{figure}

The collected object profiles amount to $5.5$ GB of disk space. We notice that for $19$ of the $46$ target methods, \pankti does not collect all the object profiles. This means that the size threshold of $200$ MB is encountered for these $19$ methods. The distribution of the sizes of the collected object profiles is illustrated in \autoref{fig:size-distribution}. From the figure, we see that most of the collected object profiles for PDFBox range from $0$ to $10$ MB in size. The smallest object profile we collect is $158$ bytes in size, with the largest one being nearly $55$ MB. Larger object profiles correspond to complex objects with many levels of nesting, the maximum level of nesting we observe in a collected object profile is $36$. As explained in \autoref{sec:selection-object-profiles}, we limit the total storage size for object profiles; to trade between the value of the collected profiles and the performance of \pankti. In the extreme cases, such as those in rows $31$, $33$, and $34$ in \autoref{tab:experiment-results}, \pankti must store only a fraction of the profiles.

\begin{lstlisting}[language=Java, belowskip=-10pt, label={lst:isFill-method}, caption={A target method called \texttt{isFill} in PDFBox}, float]
public boolean isFill() {
  return this == FILL || this == FILL_STROKE || this == FILL_CLIP || this == FILL_STROKE_CLIP;
}
\end{lstlisting}

\begin{figure}[h]
\centering
\includegraphics[width=\columnwidth]{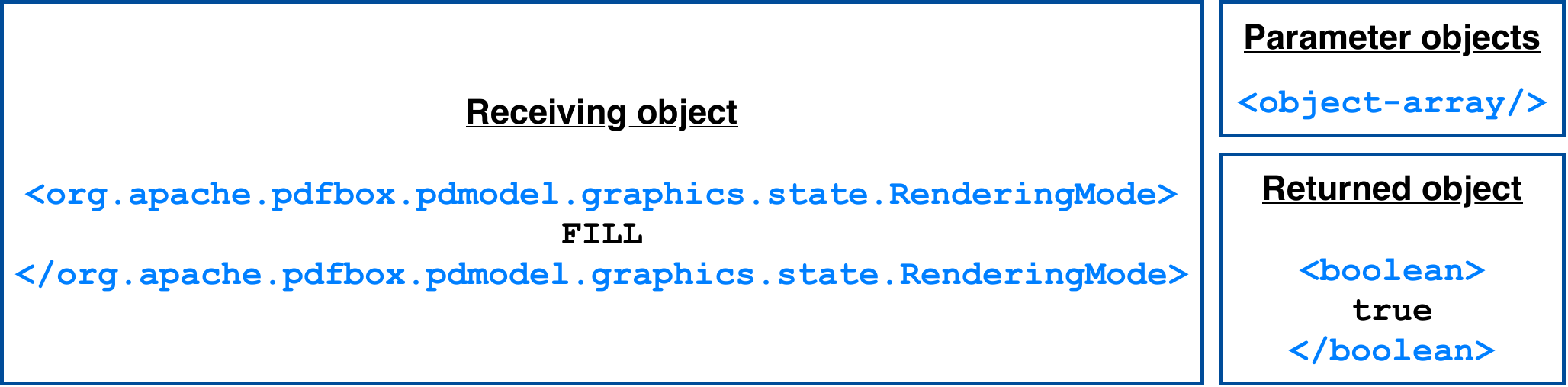}
\caption{A production object profile for the method \texttt{isFill()} in PDFBox}
\label{fig:isFill-object-profile}
\centering
\end{figure}

\begin{lstlisting}[language=Java, belowskip=-10pt, label={lst:isFill-test}, caption={A generated differential unit test for the \texttt{isFill} method}, float]
@Test
public void testIsFill() {
  String receivingObjectStr = 
  "<org.apache.pdfbox.pdmodel.graphics.state.RenderingMode>
  FILL</org.apache.pdfbox.pdmodel.graphics.state.RenderingMode>";
    
  RenderingMode receivingObject = deserializeObject(receivingObjectStr);
    
  assertEquals(true, receivingObject.isFill());
}
\end{lstlisting}

The set of $57,563$ object profiles includes $13,851$ profiles that are unique. Per the algorithm presented in \autoref{sec:test-gen}, \pankti generates one test method corresponding to each of these profiles. The number of generated tests ranges between a minimum of $1$ for 7 target methods to a maximum of $11,730$ for 1 target method. In total, we execute $13,851$ differential unit tests generated by \pankti. $13,614$ ($98.3\%$) of these tests pass while $237$ tests fail. We discuss the reasons for the failure of the generated tests in \autoref{sec:failing-tests}.

Thanks to these generated tests, $28$ of the $46$ target methods ($60.9\%$) switch from being pseudo-tested to well-tested. 
For example, the method called \texttt{isFill} in the class \texttt{RenderingMode} (row $30$ in \autoref{tab:experiment-results}) is one method that becomes well-tested after the addition of a \pankti-generated test to the test suite of PDFBox.

Let us now discuss the case of the heavily-invoked method \texttt{isFill}, shown in \autoref{lst:isFill-method}. It takes no parameter, and returns a boolean value, depending on certain values of the rendering mode for a PDF document. It is invoked $30,944$ times by our production workload. We collect all $30,944$ object profiles for this method, of $158$ bytes each. Since all these object profiles are the same, shown in \autoref{fig:isFill-object-profile}, \pankti generates one single test method for it. Its status upgrades from pseudo-tested to well-tested with the addition of this new test. The generated test is shown in \autoref{lst:isFill-test}.

We formatted the generated test to be consistent with the structure of test methods in the PDFBox codebase, and suggested it to the developers of PDFBox in the form of a pull request\footnote{\url{https://github.com/apache/pdfbox/pull/88}} on  GitHub. The pull request was accepted by a developer, and the new test case is now part of the test suite for PDFBox. This indicates that \pankti can capture relevant information in production and turn them into valid test inputs and oracles.

\begin{mdframed}[style=mpdframe,frametitle=Highlight from the PDFBox Experiment]
With a PDF manipulation workload for PDFBox, \pankti generates differential unit tests that improve the testing of $28$ out of $46$ target pseudo-tested methods. This case study shows that the \pankti monitoring and test generation pipeline scales to thousands of method invocations and object profiles. Furthermore, this case study validates the design decision of thresholding the number and size of collected object profiles. 
\end{mdframed}

\subsection{Case Study 3: Broadleaf}

Descartes finds $32$ pseudo-tested methods when running the original test suite of the Broadleaf e-commerce application. Next, we deploy Broadleaf together with \pankti and execute it with the typical e-commerce workload described in \autoref{sec:workloads}. \pankti finds that $11$ out of the $32$ pseudo-tested methods are executed by our workload. These $11$ effective target methods are shown in rows 76 to 86 in \autoref{tab:experiment-results}. They are invoked $867$ times during the experiment. The size threshold of $200$ MB is not encountered for any of these methods. We observe that target methods are invoked throughout the duration of the experiment. Consequently, the object profiles for all invocations are successfully serialized to files by \pankti. During the test generation phase, \pankti identifies $351$ unique object profiles from the collected ones. All of the unique object profiles are transformed into unit tests without any exception.

When executing the $351$ generated differential unit tests, $244$ ($69.5\%$) of them pass, while the other $107$ ($30.5\%$) fail. There are several reasons why some generated tests may fail, which will be discussed in \autoref{sec:failing-tests}. The passing tests improve the test quality for $6$ out of the $11$ ($54.5\%$) target methods. This means that \pankti generates tests that better specify the behavior of these $6$ methods.

Let us take row 80 in \autoref{tab:experiment-results} as an example. In Broadleaf's original test suite, the method \texttt{getHasOrderAdjustments} in class \texttt{OrderImpl} is pseudo-tested. If the body of this method is replaced by \texttt{return true;} or \texttt{return false;}, none of the $6$ test cases that reach it fail, meaning that even though the method is covered, its correct behavior is not specified by the existing tests.
In production, \pankti collects $6$ unique object profiles for this method. \autoref{lst:test-case-broadleaf} presents one of the test cases generated according to the object profiles. In the generated test, the receiving object is deserialized from the profile (since the profile is large, the object profile is read from a file for the sake of test readability) (line 4). According to the collected production data, the method invocation is expected to return \texttt{false}. Thus \texttt{assertEquals(false, ...)} is generated to verify the actual return value of method \texttt{getHasOrderAdjustments} (line 5). After the addition of this newly generated differential unit test to the test suite of Broadleaf, Descartes does not detect \texttt{getHasOrderAdjustments} as a pseudo-tested method anymore. This implies that the values collected by observing the interactions of a real user with the website can contribute to the improvement of the overall quality of the test suite of Broadleaf.

\begin{lstlisting}[language=Java, belowskip=-10pt, label={lst:test-case-broadleaf}, caption={A generated differential unit test for the \texttt{getHasOrderAdjustments} method}, float]
@Test
public void testGetHasOrderAdjustments() {
  File fileReceiving = new File("receiving.xml");
  OrderImpl receivingObj = deserializeObject(fileReceiving);
  assertEquals(false, receivingObj.getHasOrderAdjustments());
}
\end{lstlisting}

\begin{mdframed}[style=mpdframe,frametitle=Highlight from the Broadleaf Experiment]
\pankti collects $351$ unique object profiles while running the production workload on the e-commerce application Broadleaf, and successfully generates a test case for each profile. In total, $244$ test cases improve the test thoroughness for $6$ out of the $11$ target pseudo-tested methods. Overall, this case study shows that \pankti works well in a typical web, HTTP-based architecture which is representative of many enterprise applications.
\end{mdframed}

\subsection{Experimental Results with System Tests}
\label{sec:results-system-test}

\begin{table*}
\centering
\caption{Experimental results from executing system tests of the study subjects used to evaluate \pankti}
\label{tab:system-test-results}
\begin{tabular}{@{}lrrrrrc@{}}
\toprule
\textbf{PROJECT} & \textbf{\#SYSTEM\_TESTS} & \textbf{\#EFF\_TARGET} & \textbf{\#PANKTI\_TESTS} & \textbf{\#PASSING} & \textbf{\#FAILING} & \textbf{PANKTI\_STATUS} \\ \midrule
Jicofo & 11 & \textbf{29} / 29 & 20 & \textbf{20} / 20 & \textbf{0} / 20 & \begin{tabular}[c]{@{}c@{}}\textbf{well-tested: 18} / 29 \\ \textbf{pseudo-tested: 11} / 29 \end{tabular} \\ \midrule
PDFBox & 150 & \textbf{111} / 138 & 18,291 & \textbf{17,655} / 18,291 & \textbf{636} / 18,291 & \begin{tabular}[c]{@{}c@{}}\textbf{well-tested: 66} / 111 \\ \textbf{pseudo-tested: 45} / 111 \end{tabular} \\ \midrule
Broadleaf & 6 & \textbf{26} / 32 & 692 & \textbf{684} / 692 & \textbf{8} / 692 & \begin{tabular}[c]{@{}c@{}}\textbf{well-tested: 12} / 26 \\ \textbf{pseudo-tested: 14} / 26 \end{tabular} \\ \midrule
\textbf{TOTAL} & 167 & \textbf{166} / 199 & 19,003 & \textbf{18,359} / 19,003 & \textbf{644} / 19,003 & \begin{tabular}[c]{@{}c@{}}\textbf{well-tested: 96} / 166 \\ \textbf{pseudo-tested: 70} / 166 \end{tabular} \\
\bottomrule
\end{tabular}
\end{table*}

\autoref{tab:system-test-results} summarizes the results of the assessment of \pankti's ability to trace the execution of system tests in order to generate differential unit tests, per the protocol of \autoref{sec:system-tests}.
For each of the three study subjects. the number of system tests we identify is given in the column \#SYSTEM\_TESTS.
\#EFF\_TARGET represents the number of target pseudo-tested methods invoked during the execution of these system tests.
The number of differential unit tests generated using the object profiles collected by \pankti is signified by \#PANKTI\_TESTS.
Of these, the tests that pass are represented as \#PASSING, and the ones that do not, as \#FAILING.
In column PANKTI\_STATUS we provide the status of the invoked target methods after adding the \pankti-generated differential unit tests to the test suite of the project.
They either remain pseudo-tested, or become well-tested.

The table indicates  that there are $11$, $150$, and $6$ system tests in Jicofo, PDFBox, and Broadleaf, respectively. \pankti instruments $29$, $138$, and $32$ pseudo-tested methods in these projects. Of these instrumented methods, $29$, $111$, and $26$ are invoked during the execution of the system tests, and are the effective targets for test generation. \pankti successfully generates $20$, $18,291$, and $692$ differential unit tests for these target methods. Of these tests, $100\%$,  $96.5\%$, and $98.8\%$ pass. As a result of the tests generated by \pankti, $62\%$ ($18$ of $29$), $59.5\%$ ($66$ of $111$), and $46.1\%$ ($12$ of $26$) of targets in Jicofo, PDFBox, and Broadleaf are no longer pseudo-tested.

We note that the set of methods invoked during our experiments with production workloads differs from the ones invoked with system tests, i.e., for the same \#TARGET, \#EFF\_TARGET is different for the two experiments. We observe that the production workload for PDFBox invokes $2$ methods which are never reached by system tests. One of these methods becomes well-tested as a result of the tests generated by \pankti. For Broadleaf, the production workload reaches $3$ methods not invoked at all by system tests, of which $2$ methods become well-tested due to the addition of the new tests. For Jicofo, both the production workload and the system tests cover the  same set of methods. These observations confirm that system tests can miss some behavior exercised in production \cite{wang2017behavioral}. Consequently, a production workload is an essential complement to system tests for test improvement, as it provides additional test generation targets for \pankti.

In total, \pankti generates tests for $166$ out of the $199$ instrumented pseudo-tested methods, using $167$ system tests across the three study subjects. $96$ of these methods ($57.8\%$) become well-tested as a result of the tests automatically generated by \pankti. This experiment demonstrates that \pankti can  generate differential unit tests from system tests.

\section{Discussion}\label{sec:discussion}
We now discuss the key aspects of \pankti.

\subsection{When \pankti-generated Tests Fail}
\label{sec:failing-tests}

From \autoref{tab:experiment-results}, we observe that a total of $344$  \pankti-generated unit tests fail ($2.4\%$ of the generated tests). We manually analyze a random sample of them to understand the causes.
First, we find that no test generated for methods that return a primitive type, an array of a primitive type, a wrapper object for a primitive type, or an object of an enum type fail. 
For the tests that fail in PDFBox and Broadleaf, all methods return complex objects. Our manual analysis reveals four 
main reasons for a failure in a \pankti-generated test.

\emph{Comparison with the \texttt{equals} method can fail for arbitrary objects.} An  \texttt{assertEquals} JUnit assertion internally invokes the \texttt{equals} method on the two objects being compared. According to the default implementation of \texttt{equals}, a value of \texttt{true} is returned for two objects if they have the same reference or address in memory. To perform a deep comparison on the two objects, the behavior of \texttt{equals} must be overridden \cite{object_equality}. As this is not guaranteed for the arbitrary objects we serialize during our experiments, the equality assertions, even for two objects that are otherwise internally identical, are prone to failure if the \texttt{equals} method is not, or partially, implemented. This is the case for the target methods of Broadleaf in row 85, and PDFBox in rows 51, 72, and 73 of \autoref{tab:experiment-results}.

\emph{An external component or a cache is not available during testing.} Tests can fail due to the unavailability of external components at testing time (while being available in production). For example, the method \texttt{getName} in Broadleaf (row 77 in \autoref{tab:experiment-results}) returns the name of a product as a \texttt{java.lang.String} object. When \pankti collects object profiles in the production environment, the method \texttt{getName} returns the name of a product provided through requests to an external translation service.
However, during test execution, the method \texttt{getName} always returns the default English name of a product because the translation service is not activated. 
Tests may also fail due to the unavailability of internal caches. The target methods for PDFBox in rows 37, 44, 45, 46, and 50 in \autoref{tab:experiment-results} access objects from in-memory caches implemented using \texttt{java.lang.ref.SoftReference}, which are not available  in the testing environment, causing the assertions to fail. 

\emph{The target methods return objects containing transient fields.} The objects being returned by the target methods for PDFBox in rows 33, 34, 35, 53, and 73 of \autoref{tab:experiment-results} contain transient fields which are not serialized, by definition. Examples of such objects include \texttt{java.awt.geom.GeneralPath} and \texttt{java.util.IdentityHashMap}. This leads to failing assertions during the comparison of the returned object during the method invocation in the generated test.

\emph{The class declaring the target method overrides the serialization behavior.} XStream has different converters for different kinds of classes. If a class defines its own serialization behavior, for example, by implementing the \texttt{java.io.Externalizable} interface, XStream follows it. If the serialization or the deserialization behavior is not well designed, it can possibly lead to an assertion failure. The tests for method \texttt{getBaseRetailPrice} in Broadleaf (row 83 in \autoref{tab:experiment-results}) fail for this reason. The method returns an object whose type is \texttt{Money}. \texttt{Money} customizes its deserialization such that a \texttt{java.math.BigDecimal} field is converted from a \texttt{float} value. However, this changes the scale of the \texttt{BigDecimal}, which makes the assertion fail in the generated test.

\subsection{Cases where the Test Suite is not Improved}
From \autoref{tab:experiment-results}, we see that \pankti generates at least one differential unit test that passes for  $15$ target methods,  yet their status does not upgrade from pseudo-tested to tested. We analyze these cases manually. For those $15$ target methods, \pankti generates valid differential unit tests that check the behavior of these target methods when they return \texttt{null}, which is still a weak assertion.

We run Descartes, to determine the status of the target methods after adding the \pankti-generated tests. For each of these $15$ methods, Descartes performs one transformation on the methods: replace the body with \texttt{return null;}.
The \pankti-generated tests do not distinguish this transformation from the normal behavior of the method, and thus do not fail. 
This is a corner case of the default test adequacy criterion of \pankti based on pseudo-tested methods. \pankti generates valid, relevant test cases, yet the extreme case where Descartes generates a single variant goes unnoticed.

\subsection{Overhead of Generating Tests from Production Workloads}

This section reports a thorough performance evaluation of the deployment overhead of \pankti for the three study subjects.
For PDFBox, we compute the average CPU and memory utilization of $10$ normal executions of the workload defined in \autoref{sec:prod-workload-pdfbox}, $10$ executions of this workload which we monitor with  Glowroot only, and $10$ executions with \pankti fully attached (Glowroot, together with the instrumentation and serialization). 
We find that CPU usage is $3.7$\% during normal execution, and $21$\% with Glowroot attached. We also observe that attaching \pankti does not introduce any additional increase in CPU usage. 
Memory usage is $65$ MB during normal execution, $180$ MB while execution of the workload with Glowroot, and $557$ MB after attaching \pankti.

Overhead is an important aspect for interactive client-facing applications such as Jitsi/Jicofo (videoconference) and Broadleaf (e-commerce), because it may degrade user experience. For a Jitsi videoconference, without \pankti, we observe an average CPU and memory usage of $0.1\%$ and $338$ MB, respectively. When attaching \pankti, we observe $5.5\%$ CPU usage and $902$ MB memory.
These values drop to $0.4\%$ and $688$ MB, respectively, when ignoring the target methods that raise exceptions related to serialization (discussed in \autoref{sec:results-jicofo}). Packet loss rate remains stable at zero for all experiments, implying that users do not experience any side effects on video or audio quality from lost packets due to \pankti. For Broadleaf, the average CPU and memory usage during normal execution are $16.4\%$ and $1081$ MB, respectively, and $41.8\%$ and $1130$ MB with \pankti attached. To determine the impact on user experience, we measure the average response time for HTTP requests with \texttt{apache-utils}\footnote{\url{https://httpd.apache.org/docs/2.4/programs/ab.html}} and find that it increases by an average of 185 milliseconds. The results from these experiments can also be found at \url{https://github.com/castor-software/pankti-experiments}.

\pankti builds on top of Glowroot, a widely-adopted, state-of-the-art monitoring solution for Java projects. 
Glowroot constitutes a significant part of the overall \pankti overhead, yet, it is still acceptable for a standard server running in a real production scenario \cite{cornejo2020field}, and this computation price is paid by many applications in the world, in order to access the state-of-the-art monitoring provided by Glowroot.  It is also important to note that these values of overhead are subject to change depending on the workload, the set of target methods, and the threshold for object profile collection, all of which are completely configurable factors.

\subsection{Coverage Change with \pankti}
The code coverage ratio of the three case studies discussed in \autoref{sec:evaluation} slightly increases after adding the tests generated by \pankti to the original test suites. For Jicofo, the \pankti differential unit tests cover five new lines of code and six new branches. For PDFBox, the addition of the new tests results in the coverage of four more methods in the application. For Broadleaf, the new tests increase test suite coverage by one line and two branches.

This outcome is to be expected because the test adequacy criterion used in our evaluation is pseudo-tested methods (see \autoref{sec:targets-evaluation}). Recall that, by construction, these target methods are already covered by the original test suite of the three projects. 
For this reason, the only coverage increase may happen in a few uncovered branches in covered methods.
To this extent, code coverage can be considered as irrelevant. Instead, we fully focus on the quality of the oracle, and assess the improvement provided by \pankti tests with regards to the oracle: still pseudo-tested or not, after adding the \pankti tests. An interesting area of future work would be to study the impact of \pankti-generated tests on coverage, when the targets for test generation are also selected based on coverage.

\subsection{Privacy Implications}

As regulations, such as the General Data Protection Regulation (GDPR) in the European Union, gain traction, stricter policies are being enforced to address issues concerning the storage and processing of user data. 
One technical challenge induced by these regulations consists in finding a good balance between effective monitoring of software systems in production, and secure and lawful  handling of user data \cite{menges2021towards}. \pankti has privacy implications, since the production workload and object profiles may contain sensitive information, such as user names and passwords. We mitigate a part of this risk with one explicit input that specifies a selection of target methods to be instrumented: the list of target methods can be audited and adjusted by developers first, according to privacy and regulations. In this way, \pankti would focus its instrumentation and monitoring on non-sensitive methods only.
Without involving developers, it is also possible to protect users' privacy with the state-of-the-art of automated privacy-preservation techniques \cite{Zhang:software-usage-analysis:2020,pulls:10.1145/2517840.2517847}.
Overall, similar to related work on production monitoring, \pankti has to trade-off between its capabilities and compliance with data protection laws \cite{shah2019analyzing}.

\subsection{Structure and Readability of the \pankti Differential Unit Tests}
Readability is an important aspect of automatically generated tests \cite{7810699, daka2015modeling}. A test produced by an automated generation tool is more likely to be seamlessly integrated into an existing test suite if developers can understand it. To this end, we engineer \pankti to ensure that generated tests are targeted and focused. For our three study subjects, each of the tests generated by \pankti contains a clear test intention, represented in the form of a single invocation of the target method with a unique input, and a single assertion on equality of the output of the invocation and the oracle. This systematic template and precise intention ensure that these tests are inherently well-structured and easy to understand. The qualitative feedback obtained on the \pankti test given as pull-request (see \autoref{sec:results-pdfbox}) tends to validate this.

As described in \autoref{sec:test-gen}, the tests generated by \pankti rely on production object profiles which are deserialized from the XML format during the execution of the test. Despite the well-defined structure of the tests themselves, we appreciate that these profiles can often be lengthy and potentially challenging to comprehend. As future work, we plan to conduct user studies involving developers in order to analyze the readability and debuggability \cite{ko2004designing} of the tests generated by \pankti.

\subsection{Threats to Validity}

The main internal threat to validity comes from the libraries that \pankti uses for the implementation. In the current version, XStream is used as the serialization and deserialization library. Though XStream is the state-of-the-art tool for object serialization, it sometimes fails for some special types of objects if it is not customized. For example, objects that are based on threads or thread data can not be serialized by default\footnote{\url{https://x-stream.github.io/faq.html\#Serialization_types}}. Yet, a serialization failure rarely happens for our experiments: we observe this only once in Jicofo. The workaround is engineering: it is possible to improve the ability of XStream's serialization by registering customized object converters.

The threat to external validity is related to the breadth of considered application domains.
We mitigate the external threat to validity by evaluating \pankti with three Java projects, as discussed in \autoref{sec:evaluation}. The main strength is that they are real-world applications that cover different production workloads and diverse production environments. We look forward to applying \pankti with our industry partners to further improve external validity. 

\section{Related Work}\label{sec:related-work}

\pankti contributes to the field of automatic generation and automatic improvement of test suites. The generation of relevant test inputs \cite{anand2013orchestrated} is a key challenge in both these domains, which has been addressed through symbolic \cite{fraser2009testing}, model-based \cite{utting2012taxonomy}, or search-based techniques \cite{mcminn2011search}. The novelty of \pankti is twofold:
first, to collect data in production and automatically turn them into test inputs;
second, to target specific parts of the application code that are weakly-tested and of interest for the developers, in order to generate test cases that are valuable for the quality of the test suite.  

The closest work to \pankti is a recently developed tool called Replica, by Wang and Orso \cite{wang2020improving}. It traces production behavior as a sequence of method calls and uses these traces to spot behaviors triggered in production but not covered by the test suite. Replica then uses a guided symbolic execution of the program to ``mimic'' this behavior and to generate inputs for untested behavior. 
The primary difference with \pankti is that we generate test oracles from the actual observations made in production, and not from symbolic execution. \pankti also collects a different type of trace compared to Replica: object profiles instead of sequences of method calls. Focusing on object profiles, we can generate test data that recreate production conditions.

\subsection{Generating tests from execution traces} Thummalapenta \textit{et al.} \cite{thummalapenta2010dygen}  mine program execution traces, which include method calls, arguments, and return values, to generate parameterized unit tests implemented as regression tests with Pex \cite{tillmann2008pex}. Marchetto \textit{et al.} \cite{marchetto2008state} define a methodology to generate Selenium tests from event logs for web applications. Sampath \textit{et al.} \cite{sampath2007applying} apply incremental concept analysis to cluster similar test cases, generated from user-session based testing of web applications. 
Several works generate test cases that reproduce failures  \cite{DerakhshanfarDP20,XuanXM15,artzi2008recrash}. In particular, Artzi \textit{et al.} propose ReCrash \footnote{\url{http://groups.csail.mit.edu/pag/reCrash/}} to turn runtime observations into unit tests that reproduce failures \cite{artzi2008recrash}. ReCrash works in a similar fashion as \pankti in that it serializes object states observed during execution. However, unlike \pankti, its goal is not to generate tests for target methods that need better testing, but to generate tests that raise the same exceptions as observed in production. In particular, ReCrash has no feature to create derived oracles. More recently, Utting \textit{et al.} \cite{utting2020identifying} apply machine learning  to user and test execution traces in order to identify test inadequacies and generate new tests for usage scenarios that are missing from the test suite. Křikava and Vitek \cite{kvrikava2018tests} record function calls, including arguments, in execution traces of software packages developed in the R language, and extract unit tests from them with the goal of improving coverage.
Similar to these studies, \pankti relies on the observation of an executing application for test generation. Compared to these tools, the key novelty of \pankti is the execution representation using object profiles, which has never been proposed so far. Moreover, \pankti allows developers to choose which methods to target for test generation.

\subsection{Test suite improvement}
Danglot \textit{et al.} \cite{danglot2019automatic} and Baudry \textit{et al.} \cite{BaudryFJT02} propose to transform existing test methods into new ones that globally improve the coverage and the mutation score of the test suite \cite{DeMilloLS78}. Tillman and Schulte \cite{tillmann2006unit} suggest the use of symbolic execution to generalize traditional unit tests into parameterized unit tests and instantiating these PUTs to obtain concrete tests for higher test coverage. 
Saff \textit{et al.} \cite{saff2005automatic} develop a test refactoring technique to automatically generate unit tests from bulky system tests.
Harder \textit{et al.} \cite{harder2003improving} compute the operational difference based on the behavior of a program observed during execution in semantic terms with Daikon \cite{ernst2007daikon} to keep only those test cases that add to the abstraction. This results in a test suite that is minimal but better at detecting faults than a suite with high coverage. Chen \textit{et al.} \cite{chen2020taming} address the issues related to behavioral backward incompatibilities (BBIs), and propose the early detection of BBIs in libraries across their client projects by prioritizing and executing the clients' tests. The goal of these studies, like ours, is to improve the test quality of the project. The key differences are that \pankti 
bases its test input and derived oracle generation process on observations collected in production, and that it targets test generation on a subset of methods that the developers consider as relevant for test improvement.

\subsection{Handling the oracle problem}
The oracle problem in software testing refers to the identification of the desired behavior of a program unit \cite{StaatsWH11}. 
In their survey, Barr \textit{et al.} \cite{barr2014oracle} investigate four aspects in the literature that address the oracle problem, namely specified, derived, and implicit oracles, and the lack of automated test oracles. Gay  \textit{et al.} \cite{GaySWH15a} combine monitoring the execution of test cases, with mutation testing, in order to automatically select oracle data. Bertolino \textit{et al.} \cite{bertolino2020run} execute tests in-vivo, as an application executes, effectively leveraging production data as oracles and allowing the discovery of corner-cases that may otherwise be rare in the testing environment. The oracle can be automated, taking the domain into account. For instance, the ADVISOR tool of Genç \textit{et al.} \cite{gencc2019advisor} automates the test oracle in systems with a graphical/visual output, which can otherwise be very inaccurate due to contextual differences between an output image and a reference image. \pankti approaches the oracle problem by automatically synthesizing derived oracles from production observations, which is novel to the best our knowledge. 

Our work is different from the technique of  Elbaum \textit{et al.} \cite{elbaum2006carving}, who also generate derived oracles: they use system tests whereas \pankti uses production workloads;
their approach does not allow the set of target methods to be configured according to a developer-defined criterion. Moreover, \pankti can generate tests for methods that are never tested, while the technique of Elbaum \textit{et al.} generates test cases exclusively for methods that are covered by system-level tests

\subsection{Capture and replay}
Joshi and Orso \cite{joshi2007scarpe} present a capture and replay tool, and propose potential applications, including the generation of regression tests. Their tool captures selectively at the boundary of a subset of the application under study as it executes. Steven \textit{et al.} \cite{steven2000jrapture} design a tool called JRapture that captures the sequence of interactions between an executing Java program and components on the host system such as files, or events on graphical user interfaces. These sequences can then be replayed for observation-based testing. GenUTest \cite{pasternak2009genutest} is a capture and replay tool which logs method calls and the sequence of their occurrence, in a medium-sized, executing instrumented program. The arguments passed to and values returned from methods are serialized into logs, which are utilized for the generation of unit tests and mocks for methods, including test assertions. Saieva \textit{et al.} \cite{saieva2020ad} generate ad-hoc tests that replay recorded user execution traces in order to test candidate patches for critical security bugs.
The uniqueness of \pankti is that object profiles are captured not for the purpose of replaying the same sequence of operations on the system, but for specifying the behavior of target methods, through focused differential unit tests, with derived test oracles. This is in contrast to capture and replay systems, which typically lack oracles.

\section{Conclusion}\label{sec:conclusion}
This paper has introduced \pankti, a tool that observes Java programs in production to automatically generate differential unit tests.
\pankti observes specific methods that are weakly-tested according to a test adequacy criterion and introduces a novel technique based on the collection of object profiles. We have conducted experiments with three sizeable, popular, multi-domain,  open-source Java projects to assess \pankti's ability at monitoring production and at improving real-world test suites. \pankti successfully generates differential unit tests that improve the testing of $53$ of the $86$ ($61.6\%$) methods targeted across our three study subjects. This shows that \pankti is able to generate tests for real-world Java software.

In follow-up studies, we will broaden the scope of test improvement to other weakly-tested parts.
In particular, one can focus on methods that are not reached by any existing test but executed in the field, the existence of which has been demonstrated by Wang \textit{et al.} \cite{wang2017behavioral} and Gittens \textit{et al.} \cite{gittens2006all}. Also, we wish to explore test input minimization by utilizing partial object profiles instead of the whole. Our second thread for future work consists in extending \pankti to generate integration- or system-level tests, by considering arbitrary sequences of method invocations across different classes.

\balance
\bibliographystyle{ieeetr}
\bibliography{main}

\end{document}

%% file: table-experiment-results.tex
\begin{table*}
\centering
\tiny
\caption{\reviseadd{Experimental results from executing production workloads of the study subjects used to evaluate \pankti}}\label{tab:experiment-results}
\begin{adjustbox}{totalheight=\textheight-2\baselineskip,width=1\textwidth,center=\textwidth}
\begin{tabular}{rlrrrrrrc}
\hline
\textbf{\#} & \textbf{CLASS\_METHOD} & \textbf{\#INVOCATIONS} & \textbf{\#COLLECTED} & \textbf{\#UNIQUE} & \textbf{\#PANKTI\_TESTS} & \textbf{\#PASSING} & \textbf{\#FAILING} & \textbf{PANKTI\_STATUS} \\ \hline
\rowcolor[HTML]{E9EAEB} 
1 & JitsiMeetConfig\_isTccEnabled & 20 & \textbf{18} / 20 (90\%) & \textbf{1} / 18 (0.05\%) & 1 & \textbf{1} / 1 (100\%) & \textbf{0} / 1 (0\%) & \cellcolor[HTML]{D1F3D4} \textbf{well-tested} \\ 
2 & JitsiMeetConfig\_isOctoEnabled & 15 & \textbf{15} / 15 (100\%) & \textbf{1} / 15 (0.06\%) & 1 & \textbf{1} / 1 (100\%) & \textbf{0} / 1 (0\%) & \cellcolor[HTML]{D1F3D4} \textbf{well-tested} \\ 
\rowcolor[HTML]{E9EAEB} 
3 & JitsiMeetConfig\_getMinBitrate & 12 & \textbf{11} / 12 (91.6\%) & \textbf{1} / 11 (0.09\%) & 1 & \textbf{1} / 1 (100\%) & \textbf{0} / 1 (0\%) & \cellcolor[HTML]{D1F3D4} \textbf{well-tested} \\ 
4 & JitsiMeetConfig\_getStartBitrate & 12 & \textbf{11} / 12 (91.6\%) & \textbf{1} / 11 (0.09\%) & 1 & \textbf{1} / 1 (100\%) & \textbf{0} / 1 (0\%) & \cellcolor[HTML]{D1F3D4} \textbf{well-tested} \\  
\rowcolor[HTML]{E9EAEB} 
5 & JitsiMeetConfig\_isRembEnabled & 12 & \textbf{12} / 12 (100\%) & \textbf{1} / 12 (0.08\%) & 1 & \textbf{1} / 1 (100\%) & \textbf{0} / 1 (0\%) & \cellcolor[HTML]{D1F3D4} \textbf{well-tested} \\  
6 & JitsiMeetServices\_getJigasiDetector & 8 & \textbf{0} / 8 (0\%) & 0 & 0 & 0 & 0 & \textbf{pseudo-tested} \\
\rowcolor[HTML]{E9EAEB} 
7 & JitsiMeetServices\_getJibriDetector & 6 & \textbf{0} / 6 (0\%) & 0 & 0 & 0 & 0 & \textbf{pseudo-tested} \\
8 & JitsiMeetConfig\_getMinParticipants & 5 & \textbf{5} / 5 (100\%) & \textbf{1} / 5 (20\%) & 1 & \textbf{1} / 1 (100\%) & \textbf{0} / 1 (0\%) & \cellcolor[HTML]{D1F3D4} \textbf{well-tested} \\  
\rowcolor[HTML]{E9EAEB} 
9 & ColibriConferenceImpl\_hasJustAllocated & 4 & \textbf{0} / 4 (0\%) & 0 & 0 & 0 & 0 & \textbf{pseudo-tested} \\
10 & JitsiMeetConfig\_getEnforcedVideobridge & 4 & \textbf{4} / 4 (100\%) & \textbf{1} / 4 (25\%) & 1 & \textbf{1} / 1 (100\%) & \textbf{0} / 1 (0\%) & \textbf{pseudo-tested} \\
\rowcolor[HTML]{E9EAEB} 
11 & JitsiMeetConfig\_getOpusMaxAverageBitrate & 4 & \textbf{4} / 4 (100\%) & \textbf{1} / 4 (25\%) & 1 & \textbf{1} / 1 (100\%) & \textbf{0} / 1 (0\%) & \cellcolor[HTML]{D1F3D4} \textbf{well-tested} \\  
12 & JitsiMeetConfig\_getStartAudioMuted & 4 & \textbf{4} / 4 (100\%) & \textbf{1} / 4 (25\%) & 1 & \textbf{1} / 1 (100\%) & \textbf{0} / 1 (0\%) & \cellcolor[HTML]{D1F3D4} \textbf{well-tested} \\  
\rowcolor[HTML]{E9EAEB} 
13 & JitsiMeetConfig\_getStartVideoMuted & 4 & \textbf{4} / 4 (100\%) & \textbf{1} / 4 (25\%) & 1 & \textbf{1} / 1 (100\%) & \textbf{0} / 1 (0\%) & \cellcolor[HTML]{D1F3D4} \textbf{well-tested} \\  
14 & JitsiMeetConfig\_isRtxEnabled & 4 & \textbf{4} / 4 (100\%) & \textbf{1} / 4 (25\%) & 1 & \textbf{1} / 1 (100\%) & \textbf{0} / 1 (0\%) & \cellcolor[HTML]{D1F3D4} \textbf{well-tested} \\  
\rowcolor[HTML]{E9EAEB} 
15 & JitsiMeetConfig\_stereoEnabled & 4 & \textbf{4} / 4 (100\%) & \textbf{1} / 4 (25\%) & 1 & \textbf{1} / 1 (100\%) & \textbf{0} / 1 (0\%) & \cellcolor[HTML]{D1F3D4} \textbf{well-tested} \\  
16 & Participant\_hasDtlsSupport & 4 & \textbf{0} / 4 (0\%) & 0 & 0 & 0 & 0 & \textbf{pseudo-tested} \\
\rowcolor[HTML]{E9EAEB} 
17 & Participant\_hasIceSupport & 4 & \textbf{0} / 4 (0\%) & 0 & 0 & 0 & 0 & \textbf{pseudo-tested} \\
18 & Participant\_hasRtxSupport & 4 & \textbf{0} / 4 (0\%) & 0 & 0 & 0 & 0 & \textbf{pseudo-tested} \\
\rowcolor[HTML]{E9EAEB} 
19 & ProtocolProviderHandler\_getXmppConnection & 4 & \textbf{0} / 4 (0\%) & 0 & 0 & 0 & 0 & \textbf{pseudo-tested} \\
20 & JitsiMeetConfig\_isLipSyncEnabled & 3 & \textbf{3} / 3 (100\%) & \textbf{1} / 3 (33.3\%) & 1 & \textbf{1} / 1 (100\%) & \textbf{0} / 1 (0\%) & \cellcolor[HTML]{D1F3D4} \textbf{well-tested} \\  
\rowcolor[HTML]{E9EAEB} 
21 & JitsiMeetConfig\_useRoomAsSharedDocName & 3 & \textbf{3} / 3 (100\%) & \textbf{1} / 3 (33.3\%) & 1 & \textbf{1} / 1 (100\%) & \textbf{0} / 1 (0\%) & \cellcolor[HTML]{D1F3D4} \textbf{well-tested} \\  
22 & JitsiMeetServices\_getSipJibriDetector & 3 & \textbf{0} / 3 (0\%) & 0 & 0 & 0 & 0 & \textbf{pseudo-tested} \\
\rowcolor[HTML]{E9EAEB} 
23 & JitsiMeetConfig\_getAudioPacketDelay & 2 & \textbf{2} / 2 (100\%) & \textbf{1} / 2 (50\%) & 1 & \textbf{1} / 1 (100\%) & \textbf{0} / 1 (0\%) & \cellcolor[HTML]{D1F3D4} \textbf{well-tested} \\  
24 & JitsiMeetConfig\_getChannelLastN & 2 & \textbf{2} / 2 (100\%) & \textbf{1} / 2 (50\%) & 1 & \textbf{1} / 1 (100\%) & \textbf{0} / 1 (0\%) & \cellcolor[HTML]{D1F3D4} \textbf{well-tested} \\  
\rowcolor[HTML]{E9EAEB} 
25 & JitsiMeetConferenceImpl\$BridgeSession\_terminate & 1 & \textbf{0} / 1 (0\%) & 0 & 0 & 0 & 0 & \textbf{pseudo-tested} \\
26 & MaxPacketRateCalculator\_computeEgressPacketRatePps & 1 & \textbf{1} / 1 (100\%) & \textbf{1} / 1 (100\%) & 1 & \textbf{1} / 1 (100\%) & \textbf{0} / 1 (0\%) & \cellcolor[HTML]{D1F3D4} \textbf{well-tested} \\  
\rowcolor[HTML]{E9EAEB}
27 & MaxPacketRateCalculator\_computeIngressPacketRatePps & 1 & \textbf{1} / 1 (100\%) & \textbf{1} / 1 (100\%) & 1 & \textbf{1} / 1 (100\%) & \textbf{0} / 1 (0\%) & \cellcolor[HTML]{D1F3D4} \textbf{well-tested} \\  
28 & MaxPacketRateCalculator\_computeParticipantEgressPacketRatePps & 1 & \textbf{1} / 1 (100\%) & \textbf{1} / 1 (100\%) & 1 & \textbf{1} / 1 (100\%) & \textbf{0} / 1 (0\%) & \cellcolor[HTML]{D1F3D4} \textbf{well-tested} \\  
\rowcolor[HTML]{E9EAEB} 
29 & MaxPacketRateCalculator\_computeSenderIngressPacketRatePps & 1 & \textbf{1} / 1 (100\%) & \textbf{1} / 1 (100\%) & 1 & \textbf{1} / 1 (100\%) & \textbf{0} / 1 (0\%) & \cellcolor[HTML]{D1F3D4} \textbf{well-tested} \\ \hline
\rowcolor[HTML]{EFFBFA} 
 & {\textbf{JICOFO\_TOTAL}} & {\textbf{152}} & {\textbf{110} / 152 (72.4\%)} & {\textbf{20} / 110 (18.2\%)} & {\textbf{20}} & {\textbf{20} / 20 (100\%)} & {\textbf{0} / 20 (0\%)} & {\begin{tabular}[c]{@{}c@{}}\textbf{well-tested : 19} / 29 \\ \textbf{pseudo-tested : 10} / 29 \end{tabular}} \\ \hline
\rowcolor[HTML]{E9EAEB} 
30 & RenderingMode\_isFill & 30944 & \textbf{30944} / 30944 (100\%) & \textbf{1} / 30944 (0.00003\%) & 1 & \textbf{1} / 1 (100\%) & \textbf{0} / 1 (0\%) & \cellcolor[HTML]{D1F3D4} \textbf{well-tested} \\  
31 & PDSimpleFont\_toUnicode & 30840 & \textbf{37} / 30840 (0.001\%) & \textbf{35} / 37 (94.5\%) & 35 & \textbf{35} / 35 (100\%) & \textbf{0} / 35 (0\%) & \cellcolor[HTML]{D1F3D4} \textbf{well-tested} \\  
\rowcolor[HTML]{E9EAEB} 
32 & GlyfSimpleDescript\_getFlags  & 23471 & \textbf{23471} / 23471 (100\%) & \textbf{11730} / 23471 (49.9\%) & 11730 & \textbf{11730} / 11730 (100\%) & \textbf{0} / 11730 (0\%) & \cellcolor[HTML]{D1F3D4} \textbf{well-tested} \\  
33 & TTFGlyph2D\_getPathForCharacterCode & 15472 & \textbf{34} / 15472 (0.002\%) & \textbf{34} / 34 (100\%) & 34 & \textbf{0} / 34 (0\%) & \textbf{34} / 34 (100\%) & \textbf{pseudo-tested} \\
\rowcolor[HTML]{E9EAEB} 
34 & TTFGlyph2D\_getPathForGID & 15472 & \textbf{34} / 15472 (0.002\%) & \textbf{34} / 34 (100\%) & 34 & \textbf{0} /34 (0\%) & \textbf{34} / 34 (100\%) & \textbf{pseudo-tested} \\
35 & PDTrueTypeFont\_getPath & 782 & \textbf{36} / 782 (0.04\%) & \textbf{36} / 36 (100\%) & 36 & \textbf{0} / 36 (0\%) & \textbf{36} / 36 (100\%) & \textbf{pseudo-tested} \\
\rowcolor[HTML]{E9EAEB} 
36 & NamingTable\_getName & 650 & \textbf{563} / 650 (86.6\%) & \textbf{273} / 563 (48.5\%) & 273 & \textbf{273} / 273 (100\%) & \textbf{0} / 273 (0\%) & \cellcolor[HTML]{D1F3D4} \textbf{well-tested} \\ 
37 & DefaultResourceCache\_getFont & 546 & \textbf{29} / 546 (0.05\%) & \textbf{29} / 29 (100\%) & 29 & \textbf{5} / 29 (0.17\%) & \textbf{24} / 29 (82.7\%) & \textbf{pseudo-tested} \\
\rowcolor[HTML]{E9EAEB} 
38 & HorizontalMetricsTable\_getLeftSideBearing & 431 & \textbf{360} / 431 (83.5\%) & \textbf{360} / 360 (100\%) & 360 & \textbf{360} / 360 (100\%) & \textbf{0} / 360 (0\%) & \cellcolor[HTML]{D1F3D4} \textbf{well-tested} \\  
39 & TTFDataStream\_readUnsignedByteArray & 431 & \textbf{431} / 431 (100\%) & \textbf{431} / 431 (100\%) & 431 & \textbf{431} / 431 (100\%) & \textbf{0} / 431 (0\%) & \cellcolor[HTML]{D1F3D4} \textbf{well-tested} \\  
\rowcolor[HTML]{E9EAEB} 
40 & HorizontalMetricsTable\_getAdvanceWidth & 275 & \textbf{275} / 275 (100\%) & \textbf{174} / 275 (63.3\%) & 174 & \textbf{174} / 174 (100\%) & \textbf{0} / 174 (0\%) & \cellcolor[HTML]{D1F3D4} \textbf{well-tested} \\  
41 & TrueTypeFont\_getAdvanceWidth & 275 & \textbf{275} / 275 (100\%) & \textbf{174} / 275 (63.3\%) & 174 & \textbf{174} / 174 (100\%) & \textbf{0} / 174 (0\%) & \cellcolor[HTML]{D1F3D4} \textbf{well-tested} \\ 
\rowcolor[HTML]{E9EAEB} 
42 & PDSimpleFont\_hasExplicitWidth & 274 & \textbf{144} / 274 (52.5\%) & \textbf{117} / 144 (81.2\%) & 117 & \textbf{117} / 117 (100\%) & \textbf{0} / 117 (0\%) & \cellcolor[HTML]{D1F3D4} \textbf{well-tested} \\  
43 & PDFontDescriptor\_getFlags & 227 & \textbf{227} / 227 (100\%) & \textbf{54} / 227 (23.8\%) & 54 & \textbf{54} / 54 (100\%) & \textbf{0} / 54 (0\%) & \cellcolor[HTML]{D1F3D4} \textbf{well-tested} \\  
\rowcolor[HTML]{E9EAEB} 
44 & FontCache\_getFont & 115 & \textbf{115} / 115 (100\%) & \textbf{106} / 115 (92.2\%) & 106 & \textbf{83} / 106 (78.3\%) & \textbf{23} / 106 (21.7\%) & \textbf{pseudo-tested} \\
45 & DefaultResourceCache\_getColorSpace & 72 & \textbf{25} / 72 (34.7\%) & \textbf{23} / 25 (92\%) & 23 & \textbf{4} / 23 (17.4\%) & \textbf{19} / 23 (82.6\%) & \textbf{pseudo-tested} \\
\rowcolor[HTML]{E9EAEB} 
46 & DefaultResourceCache\_getXObject & 66 & \textbf{15} / 66 (22.7\%) & \textbf{15} / 15 (100\%) & 15 & \textbf{13} / 15 (86.6\%) & \textbf{2} / 15 (13.3\%) & \textbf{pseudo-tested} \\
47 & PDImageXObject\_getInterpolate & 66 & \textbf{8} / 66 (12.1\%) & \textbf{8} / 8 (100\%) & 8 & \textbf{8} / 8 (100\%) & \textbf{0} / 8 (0\%) & \cellcolor[HTML]{D1F3D4} \textbf{well-tested} \\  
\rowcolor[HTML]{E9EAEB} 
48 & PDImageXObject\_getMask & 66 & \textbf{13} / 66 (19.7\%) & \textbf{13} / 13 (100\%) & 13 & \textbf{13} / 13 (100\%) & \textbf{0} / 13 (0\%) & \textbf{pseudo-tested} \\
59 & PDCIDFontType2\_codeToGID & 58 & \textbf{38} / 58 (65.5\%) & \textbf{7} / 38 (18.4\%) & 7 & \textbf{7} / 7 (100\%) & \textbf{0} / 7 (0\%) & \cellcolor[HTML]{D1F3D4} \textbf{well-tested} \\  
\rowcolor[HTML]{E9EAEB} 
50 & DefaultResourceCache\_getExtGState & 49 & \textbf{24} / 49 (49\%) & \textbf{22} / 24 (91.6\%) & 22 & \textbf{4} / 22 (18.1\%) & \textbf{18} / 22 (81.8\%) & \textbf{pseudo-tested} \\
51 & PDResources\_getExtGState & 49 & \textbf{19} / 49 (38.8\%) & \textbf{18} / 19 (94.7\%) & 18 & \textbf{0} / 18 (0\%) & \textbf{18} / 18 (100\%) & \textbf{pseudo-tested} \\
\rowcolor[HTML]{E9EAEB} 
52 & PDExtendedGraphicsState\_getStrokingOverprintControl & 48 & \textbf{48} / 48 (100\%) & \textbf{3} / 48 (0.06\%) & 3 & \textbf{3} / 3 (100\%) & \textbf{0} / 3 (0\%) & \cellcolor[HTML]{D1F3D4} \textbf{well-tested} \\  
53 & COSParser\_getEncryption & 46 & \textbf{38} / 46 (82.6\%) & \textbf{33} / 38 (86.8\%) & 33 & \textbf{13} / 33 (39.4\%) & \textbf{20} / 33 (60.6\%) & \textbf{pseudo-tested} \\
\rowcolor[HTML]{E9EAEB} 
54 & XrefTrailerResolver\_getXrefType & 46 & \textbf{46} / 46 (100\%) & \textbf{17} / 46 (36.9\%) & 17 & \textbf{17} / 17 (100\%) & \textbf{0} / 17 (0\%) & \cellcolor[HTML]{D1F3D4} \textbf{well-tested} \\  
55 & PDImageXObject\_getColorKeyMask & 39 & \textbf{7} / 39 (17.9\%) & \textbf{7} / 7 (100\%) & 7 & \textbf{7} / 7 (100\%) & \textbf{0} / 7 (0\%) & \textbf{pseudo-tested} \\
\rowcolor[HTML]{E9EAEB} 
56 & PDImageXObject\_getDecode & 39 & \textbf{7} / 39 (17.9\%) & \textbf{7} / 7 (100\%) & 7 & \textbf{7} / 7 (100\%) & \textbf{0} / 7 (0\%) & \textbf{pseudo-tested} \\
57 & PDExtendedGraphicsState\_getAutomaticStrokeAdjustment & 33 & \textbf{33} / 33 (100\%) & \textbf{4} / 33 (12.1\%) & 4 & \textbf{4} / 4 (100\%) & \textbf{0} / 4 (0\%) & \cellcolor[HTML]{D1F3D4} \textbf{well-tested} \\  
\rowcolor[HTML]{E9EAEB} 
58 & PDImageXObject\_getOptionalContent & 33 & \textbf{7} / 33 (21.2\%) & \textbf{7} / 7 (100\%) & 7 & \textbf{7} / 7 (100\%) & \textbf{0} / 7 (0\%) & \textbf{pseudo-tested} \\
59 & PDExtendedGraphicsState\_getNonStrokingOverprintControl & 24 & \textbf{24} / 24 (100\%) & \textbf{3} / 24 (12.5\%) & 3 & \textbf{3} / 3 (100\%) & \textbf{0} / 3 (0\%) & \cellcolor[HTML]{D1F3D4} \textbf{well-tested} \\  
\rowcolor[HTML]{E9EAEB} 
60 & PDExtendedGraphicsState\_getOverprintMode & 24 & \textbf{24} / 24 (100\%) & \textbf{3} / 24 (12.5\%) & 3 & \textbf{3} / 3 (100\%) & \textbf{0} / 3 (0\%) & \cellcolor[HTML]{D1F3D4} \textbf{well-tested} \\  
61 & PDFontDescriptor\_isSymbolic & 24 & \textbf{24} / 24 (100\%)  & \textbf{20} / 24 (83.3\%) & 20 & \textbf{20} / 20 (100\%) & \textbf{0} / 20 (0\%) & \cellcolor[HTML]{D1F3D4} \textbf{well-tested} \\  
\rowcolor[HTML]{E9EAEB} 
62 & PDExtendedGraphicsState\_getSmoothnessTolerance & 21 & \textbf{21} / 21 (100\%) & \textbf{3} / 21 (14.3\%) & 3 & \textbf{3} / 3 (100\%) & \textbf{0} / 3 (0\%) & \cellcolor[HTML]{D1F3D4} \textbf{well-tested} \\  
63 & PDFontDescriptor\_isFixedPitch & 21 & \textbf{21} / 21 (100\%) & \textbf{6} / 21 (28.6\%) & 6 & \textbf{6} / 6 (100\%) & \textbf{0} / 6 (0\%) & \cellcolor[HTML]{D1F3D4} \textbf{well-tested} \\  
\rowcolor[HTML]{E9EAEB} 
64 & PDFontDescriptor\_isItalic & 21 & \textbf{21} / 21 (100\%) & \textbf{6} / 21 (28.6\%) & 6 & \textbf{6} / 6 (100\%) & \textbf{0} / 6 (0\%) & \cellcolor[HTML]{D1F3D4} \textbf{well-tested} \\  
65 & PDFontDescriptor\_isSerif & 21 & \textbf{21} / 21 (100\%) & \textbf{6} / 21 (28.6\%) & 6 & \textbf{6} / 6 (100\%) & \textbf{0} / 6 (0\%) & \cellcolor[HTML]{D1F3D4} \textbf{well-tested} \\  
\rowcolor[HTML]{E9EAEB} 
66 & PDExtendedGraphicsState\_getSoftMask & 20 & \textbf{20} / 20 (100\%) & \textbf{4} / 20 (20\%) & 4 & \textbf{4} / 4 (100\%) & \textbf{0} / 4 (0\%) & \textbf{pseudo-tested} \\
67 & PDICCBased\_getDefaultDecode & 16 & \textbf{16} / 16 (100\%) & \textbf{8} / 18 (44.4\%) & 8 & \textbf{8} / 8 (100\%) & \textbf{0} / 8 (0\%) & \cellcolor[HTML]{D1F3D4} \textbf{well-tested} \\  
\rowcolor[HTML]{E9EAEB} 
68 & PDCIDSystemInfo\_getOrdering & 12 & \textbf{12} / 12 (100\%) & \textbf{1} / 12 (0.08\%) & 1 & \textbf{1} / 1 (100\%) & \textbf{0} / 1 (0\%) & \cellcolor[HTML]{D1F3D4} \textbf{well-tested} \\  
69 & PDExtendedGraphicsState\_getAlphaSourceFlag & 12 & \textbf{12} / 12 (100\%) & \textbf{1} / 12 (0.08\%) & 1 & \textbf{1} / 1 (100\%) & \textbf{0} / 1 (0\%) & \cellcolor[HTML]{D1F3D4} \textbf{well-tested} \\ 
\rowcolor[HTML]{E9EAEB} 
70 & PDExtendedGraphicsState\_getNonStrokingAlphaConstant & 12 & \textbf{12} / 12 (100\%) & \textbf{1} / 12  (0.08\%) & 1 & \textbf{1} / 1 (100\%) & \textbf{0} / 1 (0\%) & \cellcolor[HTML]{D1F3D4} \textbf{well-tested} \\  
71 & PDExtendedGraphicsState\_getStrokingAlphaConstant & 12 & \textbf{12} / 12 (100\%) & \textbf{1} / 12 (0.08\%) & 1 & \textbf{1} / 1 (100\%) & \textbf{0} / 1 (0\%) & \cellcolor[HTML]{D1F3D4} \textbf{well-tested} \\  
\rowcolor[HTML]{E9EAEB} 
72 & PDDocumentCatalog\_getViewerPreferences & 8 & \textbf{8} / 8 (100\%) & \textbf{8} / 8 (100\%) & 8 & \textbf{6} / 8 (75\%) & \textbf{2} / 8 (25\%) & \textbf{pseudo-tested} \\
73 & PDCIDFontType2\_getPath & 6 & \textbf{6} / 6 (100\%) & \textbf{6} / 6 (100\%) & 6 & \textbf{0} / 6 (0\%) & \textbf{6} / 6 (100\%) & \textbf{pseudo-tested} \\
\rowcolor[HTML]{E9EAEB} 
74 & PDCIDFont\_getCIDSystemInfo & 3 & \textbf{3} / 3 (100\%) & \textbf{1} / 3 (33.3\%) & 1 & \textbf{0} / 1 (0\%) & \textbf{1} / 1 (100\%) & \textbf{pseudo-tested} \\
75 & PDCIDSystemInfo\_getRegistry & 3 & \textbf{3} / 3 (100\%) & \textbf{1} / 3 (33.3\%) & 1 & \textbf{1} / 1 (100\%) & \textbf{0} / 1 (0\%) & \cellcolor[HTML]{D1F3D4} \textbf{well-tested} \\   \hline
\rowcolor[HTML]{EFFBFA} 
& {\textbf{PDFBOX\_TOTAL}} & {\textbf{121175}} & {\textbf{57563} / 121175 (47.5\%)} & {\textbf{13851} / 57563 (24\%)} & {\textbf{13851}} & {\textbf{13614} / 13851 (98.2\%)} & {\textbf{237} / 13851 (1.8\%)} & {\begin{tabular}[c]{@{}c@{}}\textbf{well-tested : 28} / 46 \\ \textbf{pseudo-tested : 18} / 46 \end{tabular}} \\ \hline
\rowcolor[HTML]{E9EAEB} 
76 & SkuImpl\_getCurrency & 415 & \textbf{415} / 415 (100\%) & \textbf{128} / 415 (30.8\%) & 128 & \textbf{128} / 128 (100\%) & \textbf{0} / 128 (0\%) & \textbf{pseudo-tested} \\
77 & SkuImpl\_getName & 394 & \textbf{394} / 394 (100\%) & \textbf{182} / 394 (46.1\%) & 182 & \textbf{84} / 182 (46.1\%) & \textbf{98} / 182 (53.8\%) & \cellcolor[HTML]{D1F3D4} \textbf{well-tested} \\
\rowcolor[HTML]{E9EAEB}
78 & SkuImpl\_isTaxable & 24 & \textbf{24} / 24 (100\%) & \textbf{12} / 24 (50\%) & 12 & \textbf{12} / 12 (100\%) & \textbf{0} / 12 (0\%) & \cellcolor[HTML]{D1F3D4} \textbf{well-tested} \\
79 & OrderImpl\_finalizeItemPrices & 6 & \textbf{6} / 6 (100\%) & \textbf{6} / 6 (100\%) & 6 & \textbf{6} / 6 (100\%) & \textbf{0} / 6 (0\%) & \cellcolor[HTML]{D1F3D4} \textbf{well-tested} \\
\rowcolor[HTML]{E9EAEB}
80 & OrderImpl\_getHasOrderAdjustments & 6 & \textbf{6} / 6 (100\%) & \textbf{6} / 6 (100\%) & 6 & \textbf{6} / 6 (100\%) & \textbf{0} / 6 (0\%) & \cellcolor[HTML]{D1F3D4} \textbf{well-tested} \\
81& DynamicSkuPrices\_getPrice & 4 & \textbf{4} / 4 (100\%) & \textbf{2} / 4 (50\%) & 2 & \textbf{2} / 2 (100\%) & \textbf{0} / 2 (0\%) & \cellcolor[HTML]{D1F3D4} \textbf{well-tested} \\
\rowcolor[HTML]{E9EAEB}
82 & FixedPrice...Provider\_canCalculateCostForFulfillmentGroup & 4 & \textbf{4} / 4 (100\%) & \textbf{2} / 4 (50\%) & 2 & \textbf{2} / 2 (100\%) & \textbf{0} / 2 (0\%) & \cellcolor[HTML]{D1F3D4} \textbf{well-tested} \\
83 & SkuImpl\_getBaseRetailPrice & 4 & \textbf{4} / 4 (100\%) & \textbf{4} / 4 (100\%) & 4 & \textbf{0} / 4 (0\%) & \textbf{4} / 4 (100\%) & \textbf{pseudo-tested} \\
\rowcolor[HTML]{E9EAEB}
84 & SkuImpl\_getBaseSalePrice & 4 & \textbf{4} / 4 (100\%) & \textbf{4} / 4 (100\%) & 4 & \textbf{4} / 4 (100\%) & \textbf{0} / 4 (0\%) & \textbf{pseudo-tested} \\
85 & SkuImpl\_getPriceData & 4 & \textbf{4} / 4 (100\%) & \textbf{4} / 4 (100\%) & 4 & \textbf{0} / 4 (0\%) & \textbf{4} / 4 (100\%) & \textbf{pseudo-tested} \\
\rowcolor[HTML]{E9EAEB}
86 & FixedPriceFulfillmentOptionImpl\_getPrice & 2 & \textbf{2} / 2 (100\%) & \textbf{1} / 2 (50\%) & 1 & \textbf{0} / 1 (0\%) & \textbf{1} / 1 (100\%) & \textbf{pseudo-tested} \\ \hline
\rowcolor[HTML]{EFFBFA} & {\textbf{BROADLEAF\_TOTAL}} & {\textbf{867}} & {\textbf{867} / 867 (100\%)} & {\textbf{351} / 867 (40.5\%)} & {\textbf{351}} & {\textbf{244} / 351 (69.5\%)} & {\textbf{107} / 351 (30.5\%)} & {\begin{tabular}[c]{@{}c@{}}\textbf{well-tested : 6} / 11 \\ \textbf{pseudo-tested : 5} / 11 \end{tabular}}
\\ \hline
\rowcolor[HTML]{FDF1EB} 
& {\textbf{TOTAL}} & {\textbf{122194}} & {\textbf{58540} / 122194 (47.9\%)} & {\textbf{14222} / 58540 (24.3\%)} & {\textbf{14222}} & {\textbf{13878} / 14222 (97.6\%)} & {\textbf{344} / 14222 (2.4\%)} & {\begin{tabular}[c]{@{}c@{}}\textbf{well-tested : 53} / 86 \\ \textbf{pseudo-tested : 33} / 86  \end{tabular}}
\\ \hline
\end{tabular}
\end{adjustbox}
\end{table*}